\documentclass[sigconf]{acmart}

\usepackage{multirow}
\usepackage{makecell}
\usepackage{balance}
\usepackage{colortbl}
\usepackage{soul}
\usepackage{graphicx}
\usepackage{xcolor}
\usepackage{array}
\usepackage{pifont}
\usepackage{enumitem}

\definecolor{salmon}{HTML}{F8BEC0}
\newcommand{\best}[1]{%
  \begingroup\sethlcolor{salmon}\hl{#1}\endgroup%
}

\definecolor{skyblue}{HTML}{bbd0e8}
\newcommand{\second}[1]{%
  \begingroup\sethlcolor{skyblue}\hl{#1}\endgroup%
}

\definecolor{springgreen}{HTML}{d2f0ca}
\newcommand{\third}[1]{%
  \begingroup\sethlcolor{springgreen}\hl{#1}\endgroup%
}

\definecolor{rowsep}{HTML}{E6E6E6}

\newcommand{\cmark}{\ding{51}}
\newcommand{\xmark}{\ding{55}}
\newcommand{\pmark}{\ding{109}}

\AtBeginDocument{%
  }

\setcopyright{acmlicensed}
\copyrightyear{2025}
\acmYear{2025}
\acmDOI{XXXXXXX.XXXXXXX}
\acmConference[Conference acronym 'XX]{Make sure to enter the correct
  conference title from your rights confirmation email}{June 03--05,
  2025}{Woodstock, NY}

\acmISBN{978-1-4503-XXXX-X/2025/06}

\acmSubmissionID{15}

\begin{document}
\title{A Hierarchical Quantized Tokenization Framework for Task-Adaptive Graph Representation Learning }

\author{Yang Xiang}
\orcid{0009-0007-9693-7963}
\authornote{Both authors contributed equally to this research.}
\affiliation{%
  \institution{University of Liverpool}
  \city{Liverpool}
  \state{England}
  \country{United Kingdom}
}
\affiliation{%
  \institution{Xi'an Jiaotong-Liverpool University}
  \city{Suzhou}
  \state{Jiangsu}
  \country{China}
}
\email{Y.Xiang17@liverpool.ac.uk}

\author{Li Fan}
\orcid{0009-0005-7234-5774}
\authornotemark[1]
\affiliation{%
  \institution{Hong Kong University of Science and Technology (Guangzhou)}
  \city{Guangzhou}
  \state{Guangdong}
  \country{China}
}
\email{lfan253@connect.hkust-gz.edu.cn}

\author{Chenke Yin}
\orcid{0009-0009-2176-3967}
\affiliation{%
  \institution{Jiangsu Province Science and Technology Information Institute}
  \city{Nanjing}
  \state{Jiangsu}
  \country{China}
}
\email{ckyinphd@gmail.com}

\author{Lutz Oettershagen}
\orcid{0000-0002-2526-8762}
\affiliation{%
  \institution{University of Liverpool}
  \city{Liverpool}
  \state{England}
  \country{United Kingdom}
}
\email{Lutz.Oettershagen@liverpool.ac.uk}

\author{Chengtao Ji}
\orcid{0000-0001-5733-6881}
\authornote{Corresponding author.}
\affiliation{%
  \institution{Xi'an Jiaotong-Liverpool University}
  \city{Suzhou}
  \state{Jiangsu}
  \country{China}
}
\email{Chengtao.Ji@xjtlu.edu.cn}

\renewcommand{\shortauthors}{Yang and Li et al.}

\begin{abstract}
Foundation models in language and vision benefit from a unified discrete token interface that converts raw inputs into sequences for scalable pre-training and inference. For graphs, an effective tokenizer should yield reusable discrete codes that capture both node semantics and relational structure across scales, yet prior quantization-based graph tokenizers typically combine residual vector quantization (RVQ) levels with fixed rules and often focus on a single structural view, limiting cross-task transfer. We present a hierarchical quantized tokenization framework with task-conditioned routing and dual-view token streams. It produces multi-scale codes and two synchronized sequences: a local stream that preserves node-level information and a diffusion-style multi-hop stream that summarizes connectivity. A lightweight router learns task-dependent mixtures over RVQ depths to select an appropriate granularity, while a gated cross-attention module aligns and fuses the two streams into a single token sequence without altering the downstream backbone encoder. Experiments on node classification and link prediction show consistent gains over strong quantized baselines at matched compute, with ablations verifying contributions from hierarchical quantization, adaptive routing, and fusion.
\end{abstract}



\maketitle

\section{Introduction}

Foundation Models in language and vision have scaled through an efficient discrete token interface (subword units for text and regular patches for images) that converts raw data into sequences for pre-training and inference \cite{devlin-etal-2019-bert, NEURIPS2020_1457c0d6, dosovitskiy2021an, zeng2025numina, zhang2025llm}. This interface supports transfer across datasets and tasks, enabling efficient training and inference. Extending this paradigm to graph-structured data is important but nontrivial. Graphs, which represent entities and their relations, arise in many high-impact domains, including social networks, recommender systems, biological and molecular systems, knowledge graphs, and healthcare applications \cite{scarselli2008graph,wu2020comprehensive,bronstein2021geometric,fan2019graph,hou2025heterogeneous, xiang2025harnessing, yin2025uncertainty, fan2025dual, hu2025ids, guan2025lihai}. Unlike sequences and grids, graphs are non-Euclidean, irregular, and multi-scale: local neighborhoods often capture semantic similarity, while multi-hop paths and communities capture connectivity and global structure. These properties motivate Graph Foundation Models (GFMs) \cite{liu2025graph, wang2025graph, jin2024large, li2023survey}. However, graphs still lack a unified tokenization interface that exposes multi-scale structure as reusable discrete tokens and allows downstream tasks to select which structural scales to use.

\begin{figure}[t]
  \centering
  \resizebox{0.95\linewidth}{!}{
    \includegraphics{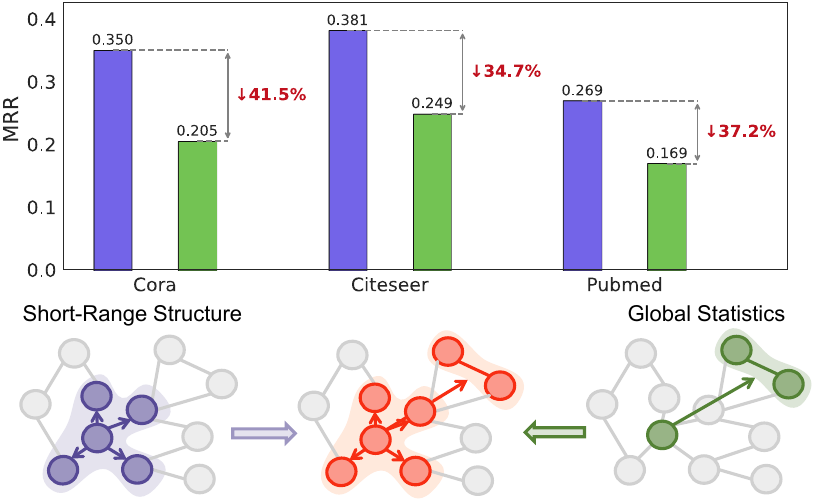}
  }
  \caption{Subfigure (a) (upper half) shows that freezing a tokenizer trained for node classification (NC) and then fine-tuning the rest of the model for link prediction (LP) leads to a large drop in LP performance. Subfigure (b) (lower half) illustrates our idea of fusing local and global information during tokenization to obtain a unified tokenizer that generalizes across tasks.}
  \label{fig:intro}
\end{figure}

Although quantization-based methods offer an efficient discrete interface for GFMs, current graph tokenizers still have two main limitations. First, graphs have a multi-scale structure, and many RVQ-based tokenizers construct hierarchical codebooks at multiple levels \cite{chen2024hight, wang2025learning}. However, when forming the final embedding, these multi-level codebooks are often combined using a simple rule such as concatenation. This makes each level contribute equally. Since the most informative scale can change with the dataset and the task \cite{xu2018representation, abu2019mixhop}, a fixed mixing strategy can underweight the levels that carry the key information and overweight useless levels that may add noise.
Second, cross-task reuse can break down when the token set is designed for only one structural level in a task, which leads to a mismatch when transferring between node-level and edge-level tasks. Using GQT \cite{wang2025learning} and LPFormer \cite{shomer2024lpformer} as examples, Figure \ref{fig:intro} compares two link prediction training protocols. The purple bar trains the model for link prediction with the tokenizer updated, while the green bar reuses a tokenizer trained for node classification and keeps it frozen during link prediction fine-tuning. The results show a substantial performance drop, suggesting that tokenizers learned for node-level objectives do not transfer well to edge-level tasks without adaptation. These findings suggest that task-specific tokenization can bias tokens toward a single structural level, which limits reuse across node-level and edge-level tasks. 

To address the limitations discussed above, we introduce Task-Adaptive Routing for Unified Graph Tokenization (TAU), a task-adaptive graph tokenizer. TAU adopts residual vector quantization with hierarchical extensions to produce multi-scale discrete tokens, and it generates two complementary token streams: a local stream that preserves node semantics and a multi-hop stream that captures connectivity. Before forming the final embedding, TAU applies lightweight task-adaptive routing over the RVQ depth to weight coarse structure relative to fine residual detail. It then uses gated cross-attention to align and fuse the two streams into a single token sequence, while keeping the backbone encoder unchanged. 

\smallskip
\noindent
\textbf{Our main contributions are:}
\begin{itemize}[leftmargin=*]
\item We present \textbf{TAU}, the first task-adaptive quantized graph tokenizer that provides a unified discrete token interface for different graph tasks. TAU generates multi-scale tokens from local and diffusion/multi-hop graph views, and adaptively routes and fuses them according to the target task.
\item We design residual vector quantization with hierarchical extensions and gated cross-attention between views, enabling parameter efficient adaptation without modifying the GFM backbone.
\item We provide ablation studies and evaluations of node classification and link prediction that isolate the effects of hierarchical quantization, routing, and fusion, demonstrating gains over strong baselines with matched compute.
\end{itemize}

\section{Related Work}
\subsubsection*{\textbf{Graph Tokenization}}
Graph tokenization is often grouped into three lines: linearized \cite{kim2022pure, chen2022nagphormer, galkin2021nodepiece}, continuous \cite{hou2022graphmae, hou2023graphmae2, thakoor2021large, velivckovic2018deep, you2020graph}, and quantization-based \cite{ding2021vqgnn, yang2024vqgraph, luo2024nid, xia2023molebert, zhuang2023imold, wang2025learning}. Linearized methods convert graphs into sequences for Transformers, such as TokenGT \cite{kim2022pure} and NAGphormer \cite{chen2022nagphormer}, and NodePiece \cite{galkin2021nodepiece} for knowledge graphs. They offer a direct language-style interface but often yield long, redundant sequences and weak global structure modeling. Continuous methods (for example, GraphMAE/GraphMAE2 \cite{hou2022graphmae,hou2023graphmae2}, BGRL \cite{thakoor2021large}, DGI \cite{velivckovic2018deep}, and GraphCL \cite{you2020graph}) learn implicit token embeddings with masked or contrastive objectives. Still, they lack discrete, composable units for a unified token interface. Quantization-based methods discretize embeddings into codebook indices. VQ-GNN \cite{ding2021vqgnn} and VQ-Graph \cite{yang2024vqgraph} reduce redundancy by reusing recurring substructures. In contrast, RVQ-based models, including NID \cite{luo2024nid}, Mole-BERT \cite{xia2023molebert}, iMoLD \cite{zhuang2023imold}, and GQT \cite{wang2025learning}, learn hierarchical discrete tokens and are mainly evaluated on node-level tasks.

\subsubsection*{\textbf{Graph Foundation Models}}
Graph foundation models extend foundation-model transfer to graph data \cite{wang2025graph} and can be viewed as task-specific or universal. Task-specific GFMs build graph Transformers with structural bias: GAT \cite{velickovic2017graph} masks attention by connectivity, and later works \cite{dwivedi2020generalization, ying2021transformers, kreuzer2021rethinking, rampavsek2022recipe, chen2022structure, wu2023sgformer, liu2023gapformer, shirzad2023exphormer, ma2023graph, deng2024polynormer} add cues like shortest-path distance and edge features. Universal GFMs aim to generalize across domains and tasks, including data-flow–graph models like GraphCodeBERT \cite{guo2021graphcodebert}, synthetic-pretraining approaches like GPT-GNN \cite{qiu2023gptgnn}, and API-chain–driven natural-language interfaces like ChatGraph \cite{peng2024chatgraph}. These studies highlight the importance of a stable graph-to-token interface for scalable pre-training and transfer.

\begin{figure*}[t]
  \centering
  \includegraphics[width=0.95\linewidth]{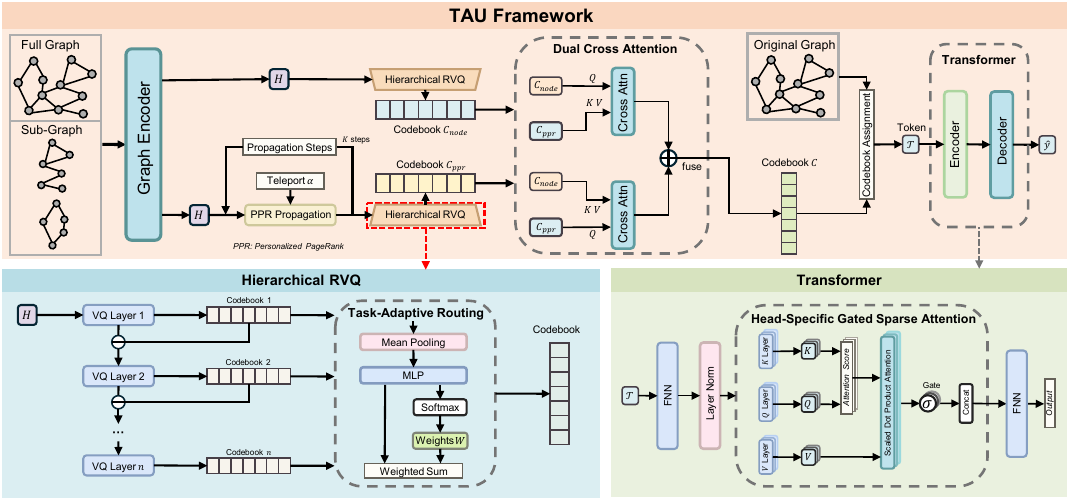}
  \caption{\textbf{Our TAU framework.} TAU produces task-unified discrete graph tokens by hierarchically quantizing both the original representations and the PPR-enhanced representations. It then uses TAQR to adaptively weight tokens at different levels. Finally, it fuses the weighted tokens into a final token sequence for the GFM backbone. The top panel shows the pipeline; the bottom panels detail the hierarchical RVQ with TAQR (left) and the transformer backbone with gated sparse attention (right).}
  \label{fig:framework}
\end{figure*}

\section{Preliminary}

We denote an input attributed graph as $G=(V,E,\mathbf{X})$, where $V$ is the set of nodes, $E\subseteq V\times V$ is the set of edges, and $\mathbf{X}\in\mathbb{R}^{n\times d}$ is the node feature matrix with $n=|V|$ nodes and feature dimension $d$. For each node $v\in V$, its input feature is denoted by $\mathbf{x}_v\in\mathbb{R}^{d}$, corresponding to the $v$-th row of $\mathbf{X}$. Let $\mathbf{h}_v^{l}$ be the node representation of node $v$ at layer $l$, with $\mathbf{h}_v^{0}=\mathbf{x}_v$. After $L$ layers, we denote the final node embedding by $\mathbf{H}\in\mathbb{R}^{n\times d_h}$, whose $v$-th row is $\mathbf{h}_v^{L}$. We use $\mathbf{A}\in\mathbb{R}^{n\times n}$ to denote the adjacency matrix and $\mathbf{D}\in\mathbb{R}^{n\times n}$ to denote the degree matrix. For residual vector quantization (RVQ), let $M$ be the number of quantization levels. The level-$m$ codebook is denoted by $\mathbf{C}^{m}\in\mathbb{R}^{k_m\times d_h}$, where $k_m$ is the number of codewords at level $m$. For node $v$, $\mathbf{q}_v^{m}$ denotes the selected codeword at level $m$, and $\mathbf{r}_v^{m}$ denotes the residual representation after the level-$m$ quantization step. We write $\mathcal{C}=\{\mathbf{C}^{1},\ldots,\mathbf{C}^{M}\}$ for the collection of all RVQ codebooks.

\section{Overall Framework}
As shown in Figure \ref{fig:framework}, TAU has two main components: (1) a task-unified hierarchical tokenizer that produces multi-scale discrete tokens and adjusts their contributions based on the task; and (2) a graph foundation model (GFM) backbone that consumes the weighted tokens to solve downstream tasks. To make the tokenizer reusable across downstream tasks, we design a task-unified hierarchical tokenizer that produces multi-level discrete tokens and adjusts their contributions based on the task. The tokenizer first encodes the input graph using a graph encoder to obtain node embeddings and Personalized PageRank (PPR)-enhanced features. It then applies hierarchical residual vector quantization to map these continuous representations into multi-level discrete tokens. To adapt quantization to different tasks and datasets, a task-adaptive quantization routing (TAQR) module learns to weight contributions from different RVQ levels when forming the final token representation. Next, a dual cross-attention module fuses tokens from the original features and the PPR-enhanced features to capture complementary information. Finally, a gating module assigns task-specific weights to token levels from different codebooks and combines them into a final token representation. The resulting weighted tokens are passed to the GFM backbone, which can be a pre-trained transformer or a GNN, to solve downstream tasks such as node classification and link prediction.

\section{Task-Unified Hierarchical Tokenizer}
In this section, we introduce our proposed task-unified hierarchical tokenizer. We use hierarchical RVQ to obtain multi-level discrete tokens from both local and PPR-enhanced features. A routing module learns task-dependent weights over RVQ levels, and a gated dual cross-attention module fuses local and global tokens into the final representation.

\subsection{Graph Encoder}
\subsubsection{Graph Encoding}
To capture both structural and semantic information from the input graph, we use a graph encoder based on graph convolutional networks (GCNs) \cite{kipf2016gcn}. The encoder takes an input graph $G=(V,E)$ with node features $\mathbf{X} \in \mathbb{R}^{n \times d}$, where $n$ is the number of nodes and $d$ is the feature dimension. We initialize the node embeddings as $\mathbf{H}^0=\{\mathbf{h}_v^0 \mid v \in V\}$ with $\mathbf{h}_v^0=\mathbf{x}_v$, and update them by applying $L$ layers of GCN:
\begin{eqnarray}\label{eq:gnn}
  \begin{aligned}
    \mathbf{h}_v^l=f_\theta^l\!\left(\mathbf{h}_v^{l-1},\; g_\phi^l\!\left(\left\{\left(\mathbf{h}_v^{l-1},\mathbf{h}_u^{l-1},\mathbf{e}_{uv}\right)\mid u\in\mathcal{N}_v\right\}\right)\right),
  \end{aligned}
\end{eqnarray}
For large-scale graphs, we adopt mini-batch training with neighbor sampling \cite{hamilton2017inductive} to efficiently compute embeddings. After $L$ encoder layers, we obtain the final node embeddings $\mathbf{H}=\{\mathbf{h}_v^L \mid v \in V\}$.

\subsubsection{PPR Propagation}
Although GCNs capture local neighborhood information effectively, our goal is to produce multi-scale discrete tokens that encode both node-level and edge-level information. PPR is effective for modeling multi-scale graph structure \cite{page1999pagerank, gasteiger2018predict, shomer2024lpformer}, and it is often beneficial for edge-level tasks. We therefore incorporate PPR-based propagation into the encoder to strengthen its multi-scale representations. Specifically, we define the PPR matrix $\mathbf{P} \in \mathbb{R}^{n \times n}$ as
\begin{eqnarray}\label{eq:ppr}
  \begin{aligned}
    \mathbf{P}=\alpha(\mathbf{I}-(1-\alpha)\mathbf{D}^{-1/2}\mathbf{A}\mathbf{D}^{-1/2})^{-1},
  \end{aligned}
\end{eqnarray}
where $\alpha$ is the teleport probability, $\mathbf{I}$ is the identity matrix, $\mathbf{A}$ is the adjacency matrix, and $\mathbf{D}$ is the degree matrix. The parameter $\alpha$ controls the balance between local and global information: a larger $\alpha$ emphasizes local structure, while a smaller $\alpha$ propagates information more broadly. Given the node embeddings $\mathbf{H}$ from the GCN encoder, we compute PPR-enhanced features by diffusion:
\begin{eqnarray}\label{eq:ppr_prop}
  \begin{aligned}
    \mathbf{H}^{(k)}=\alpha \mathbf{H}+(1-\alpha)\mathbf{D}^{-1/2}\mathbf{A}\mathbf{D}^{-1/2}\mathbf{H}^{(k-1)},
  \end{aligned}
\end{eqnarray}
for $k=1,2,\ldots,K$. We then set $\mathbf{H}_{PPR}=\mathbf{H}^{(K)}$. This iterative scheme provides a stable and efficient approximation to the closed-form diffusion in Equation \ref{eq:ppr}.

\subsection{Hierarchical Residual Vector Quantization}
\subsubsection{Residual Vector Quantization}
Following prior work \cite{wang2025learning}, residual vector quantization (RVQ) \cite{lee2022autoregressive} is effective for producing compact discrete tokens while retaining important information. We adopt RVQ to quantize both the node embeddings $\mathbf{H}$ and the PPR-enhanced features $\mathbf{H}_{PPR}$ into discrete tokens. Compared with continuous representations, vector quantization uses a small codebook and simple table lookups, reducing memory and computation requirements for large graphs. RVQ further improves expressiveness by recursively quantizing residuals from previous steps, yielding a hierarchical representation. Formally, we define an $M$-level codebook $C=\{\mathbf{c}^1, \mathbf{c}^2, \ldots, \mathbf{c}^M\}$, where each level $m$ has a codebook $\mathbf{c}^m \in \mathbb{R}^{k_m \times d_h}$ with $k_m$ codewords. For a node embedding $\mathbf{h}_v$, the RVQ procedure is
\begin{eqnarray}\label{eq:rvq}
\begin{gathered}
\mathbf{r}_v^0 = \mathbf{h}_v,\quad
\mathbf{q}_v^m = \arg\min_{\mathbf{c}^m_j \in \mathbf{c}^m}\left\lVert \mathbf{r}_v^{m-1} - \mathbf{c}^m_j \right\rVert_2,\\
\mathbf{r}_v^m = \mathbf{r}_v^{m-1} - \mathbf{q}_v^m,
\end{gathered}
\end{eqnarray}
where $\mathbf{q}_v^m$ is the selected codeword at level $m$ and $\mathbf{r}_v^m$ is the residual after level-$m$ quantization.

\subsubsection{Task-Adaptive Quantization Routing (TAQR)}
Prior tokenizers concatenate codewords from all levels equally \cite{luo2025node, wang2025learning}. This fixed mixing strategy can underweight the levels that carry the key information for a given dataset and task. Concretely, given the multi-level quantized codebooks $\mathbf{C} = \{\mathbf{c}^1, \mathbf{c}^2, \ldots, \mathbf{c}^M \}$, we aggregate them into a single token representation via task-adaptive routing over RVQ depth. Specifically, we assign a scalar routing weight to each depth. For depth $m$, we first summarize the quantized outputs by mean pooling to obtain $\mathbf{s}^m \in \mathbb{R}^{d_h}$. A two-layer MLP maps $\mathbf{s}^m$ to a logit $z^{(m)}$, and we normalize all logits with a temperature-controlled softmax:
\begin{eqnarray}\label{eq:gate}
  \begin{aligned}
    w^{(m)}=\frac{\exp \left(z^{(m)} / \tau\right)}{\sum_{j=1}^M \exp \left(z^{(j)} / \tau\right)},
  \end{aligned}
\end{eqnarray}
where $\tau>0$ controls how sharp the routing distribution is. Finally, we produce the token representation by a mixture over depths:
\begin{eqnarray}\label{eq:final_token}
  \begin{aligned}
    \mathbf{C}=\sum_{m=1}^M w^{(m)} \mathbf{C}^{(m)}.
  \end{aligned}
\end{eqnarray}
TAQR can be viewed as soft routing over quantization depth: it lets the tokenizer emphasize coarse RVQ codes or delicate residual refinements as needed by the downstream objective, without changing the backbone architecture or introducing a separate task-specific token interface. In addition, the learned weights provide a direct, interpretable signal of which quantization depths are most useful for a given training objective.

\subsection{Dual Cross Attention}
Prior work \cite{xia2024opengraph, wang2025learning} often relies on a single information source to guide tokenization, which can bias the tokenizer toward only one aspect of graph structure and hurt performance on some tasks. Although PPR propagation captures multi-scale structural signals, directly using the PPR-enhanced features $\mathbf{H}_{PPR}$ for tokenization can introduce task-irrelevant information. This issue is more visible for node-level tasks, which often depend more on local neighborhood information. To address this, we propose a dual cross-attention mechanism that jointly leverages the original node embeddings $\mathbf{H}$ and the PPR-enhanced features $\mathbf{H}_{PPR}$ during tokenization. This design aims to retain local semantics from $\mathbf{H}$ while also incorporating longer-range structure from $\mathbf{H}_{PPR}$, which is essential for supporting both node-level and edge-level tasks under a unified interface. We first apply hierarchical residual vector quantization to obtain discrete tokens from both feature sets, denoted as $\mathbf{C}_{node}$ (from $\mathbf{H}$) and $\mathbf{C}_{PPR}$ (from $\mathbf{H}_{PPR}$). We then fuse these tokens using cross attention. A single cross-attention direction implicitly assumes that one token set is always the query and the other is always the context, which may not be optimal across tasks. We therefore use dual cross attention, where each token set attends to the other:
\begin{eqnarray}\label{eq:cross_attention}
  \begin{aligned}
    \mathbf{A}_{node \rightarrow PPR} &= \text{softmax}\left(\frac{\mathbf{Q}_{node} \mathbf{K}_{PPR}^T}{\sqrt{d_h}}\right) \mathbf{V}_{PPR}, \\
    \mathbf{A}_{PPR \rightarrow node} &= \text{softmax}\left(\frac{\mathbf{Q}_{PPR} \mathbf{K}_{node}^T}{\sqrt{d_h}}\right) \mathbf{V}_{node},
  \end{aligned}
\end{eqnarray}
where $\mathbf{Q}_{node} = \mathbf{C}_{node}\mathbf{W}_Q$, $\mathbf{K}_{node} = \mathbf{C}_{node}\mathbf{W}_K$, $\mathbf{V}_{node} = \mathbf{C}_{node}\mathbf{W}_V$, and $\mathbf{Q}_{PPR}$, $\mathbf{K}_{PPR}$, $\mathbf{V}_{PPR}$ are defined analogously for $\mathbf{C}_{PPR}$. Here, $\mathbf{W}_Q$, $\mathbf{W}_K$, and $\mathbf{W}_V$ are learnable projection matrices. To suppress task-irrelevant information and allow task-dependent fusion, we add a gating mechanism that weights the two cross-attention outputs. The final fused token representation $\mathbf{C}$ is computed as
\begin{eqnarray}\label{eq:final_cross_token}
  \begin{aligned}
    \mathbf{g} &= \sigma\left(\mathbf{W}_g [\mathbf{A}_{node \rightarrow PPR}; \mathbf{A}_{PPR \rightarrow node}] + \mathbf{b}_g\right), \\
    \mathbf{C} &= \mathbf{g} \odot \mathbf{A}_{node \rightarrow PPR} + (1 - \mathbf{g}) \odot \mathbf{A}_{PPR \rightarrow node},
  \end{aligned}
\end{eqnarray}
where $\sigma(\cdot)$ is the sigmoid function, $[\cdot;\cdot]$ denotes concatenation, $\odot$ denotes element-wise multiplication, and $\mathbf{W}_g$ and $\mathbf{b}_g$ are learnable parameters. This gated dual cross-attention module enables the tokenizer to balance local and multi-hop structural signals based on the task, which improves robustness across unified graph tasks.

\section{Graph Foundation Model}
\subsection{Codebook Assignment}
After the dual cross-attention module outputs the fused codebook $\mathbf{C}$, we convert the input graph into a sequence of discrete token identifiers for the GFM. For each node $v\in V$, we assign its embedding $\mathbf{h}_v$ to a codeword in $\mathbf{C}$ and use the selected codeword index as the token id, which gives a unified token sequence for downstream backbones. For the assignment, we project node embeddings and codewords into a shared space and compute their similarities. We then obtain a codeword distribution with a temperature-controlled softmax. During training, we use the Gumbel-Softmax estimator \cite{jang2016categorical} to approximate discrete sampling and enable end-to-end learning. During inference, we use a hard assignment by selecting the most likely codeword index.

Formally, let $f_{proj}(\cdot)$ and $g_{proj}(\cdot)$ be projection layers for node embeddings and codewords. We compute similarity logits
\begin{eqnarray}
    s_{v,j} = \operatorname{sim}\bigl(f_{proj}(\mathbf{h}_v),\, g_{proj}(c_j)\bigr),
\end{eqnarray}
and the codeword distribution $\mathbf{p}_v = \mathrm{softmax}(s_{v,:}/\tau)$ where $\tau$ controls sharpness. The token sequence $\mathcal{T}=\{t_v \mid v \in V\}$ is obtained by sampling with Gumbel-Softmax during training or taking $\arg\max_j p_{v,j}$ during inference, and is fed to the GFM. This assignment scales well because it compares each node only to a fixed-size codebook, independent of graph size.

\subsection{Graph Transformer}
Given the token sequence $\mathcal{T}$ from codebook assignment, we use a Graph Transformer to solve downstream tasks by modeling dependencies among tokens. We adopt a multi-layer Transformer \cite{vaswani2017attention} with self-attention (SA) and feed-forward networks (FNN), producing token embeddings $\mathbf{Z}=\{\mathbf{z}_t \mid t\in\mathcal{T}\}$, where each $\mathbf{z}_t$ corresponds to a node. Let $\mathbf{z}_t^{(l)}$ be the representation at layer $l$. A transformer block can be defined as:
\begin{eqnarray}\label{eq:transformer}
  \begin{aligned}
    \mathbf{z}_t^{(l)} &= \text{LN}\left(\text{SA}\left(\text{LN}\left(\mathbf{z}_t^{(l-1)}\right)\right)+\mathbf{z}_t^{(l-1)}\right), \\
    \mathbf{z}_t^{(l)} &= \mathbf{z}_t^{(l)} + \text{FNN}\left(\mathbf{z}_t^{(l)}\right).
  \end{aligned}
\end{eqnarray}

To support task-unified learning, we replace standard multi-head attention with head-specific gated sparse attention \cite{qiu2025gated}. This is motivated by known limits of standard attention, including low-rank constraints \cite{srinadh2020lowrank}, weak suppression of irrelevant tokens under softmax \cite{martins2016softmax}, and attention sink effects \cite{xiao2024efficient}. In unified settings, $\mathcal{T}$ can contain many structural fragments that are not useful for a given task; per-head gating increases selectivity, improves robustness, and enhances training stability. For head $h$, we compute
\begin{equation}\label{eq:gated_attention}
\alpha_{i j}^h=\frac{\exp \left(s_{i j}^h\right)}{\sum_{i^{\prime}:\left(i^{\prime} \rightarrow j\right) \in \mathcal{E}} \exp \left(s_{i^{\prime} j}^h\right)},
\end{equation}
\begin{equation}
g_j^h=\sigma\left(q_j^h\right)\in(0,1),\qquad
o_j^h=g_j^h\sum_{i:(i \rightarrow j) \in \mathcal{E}} \alpha_{i j}^h v_i^h,
\end{equation}
where $q_j^h$, $k_i^h$, and $v_i^h$ are query, key, and value vectors, and $b_{ij}$ encodes structure between tokens $i$ and $j$ in the biased scaled dot-product score $s_{ij}^h$ (with temperature $\tau$). The head outputs $\{o_j^h\}_{h=1}^H$ are concatenated and projected by $W_o$ to form the final output. Full details are provided in Appendix \ref{appendix:gated_attention_details}.

\subsection{Lightweight Task Adapters}
After the Graph Transformer outputs shared node representations $\{\mathbf{z}_v\}$, we attach lightweight task adapters for downstream prediction. The key choice is to reuse the same backbone features for both node-level and link-level tasks, while only adding small task-specific modules. This supports unified multi-task training and evaluation without modifying the backbone.

For node classification, we use an output projection with a task-specific activation: softmax for multi-class and sigmoid for multi-label settings,
\begin{eqnarray}
    \hat{\mathbf{y}}_v^\text{NC} = h_\text{node}(\mathbf{z}_v)= \sigma_\text{out}\bigl(W_\text{out} \mathbf{z}_v + b_\text{out}\bigr),
\end{eqnarray}
and train with Negative Log-Likelihood (NLL) loss for multi-class or Binary Cross-Entropy (BCE) loss for multi-label tasks.

For link prediction, we score node pairs using a lightweight edge adapter applied to the same backbone outputs. Given $(u,v)$, we build an edge feature by combining the two node embeddings and their element-wise interaction, then predict the link probability:
\begin{eqnarray}
    \hat{y}_{uv}^\text{LP} = h_\text{edge}\bigl([\mathbf{z}_u; \mathbf{z}_v; \mathbf{z}_u \odot \mathbf{z}_v]\bigr),
\end{eqnarray}
followed by a sigmoid activation. The model is trained with BCE loss. 
Since both adapters operate on $\mathbf{z}_v$ and add a few parameters, they provide a unified interface for node and link tasks and can be swapped without changing the backbone.

\section{Experiments}
We conduct experiments to evaluate the proposed task-adaptive graph tokenizer and address the following research questions:
\begin{itemize}[leftmargin=*]
  \item \textbf{RQ1:} Can the proposed tokenizer consistently improve performance on both node classification and link prediction while remaining scalable to large graphs?
  \item \textbf{RQ2:} Can each component of the proposed tokenizer contribute to performance gains?
  \item \textbf{RQ3:} Can TAQR provide insights into which quantization depths are most useful for different datasets and tasks?
  \item \textbf{RQ4:} Can the learned codebooks be reused across tasks, indicating strong generalization?
  \item \textbf{RQ5:} How does TAQR affect the tokenizer's computational efficiency?
\end{itemize}

\subsection{Experimental Setup}
In this section, we describe the basic experimental setup, including datasets, baselines, implementation details, and evaluation metrics. The detailed experimental configurations are provided in the Appendix \ref{appendix:exp_setup}.

\subsubsection*{\textbf{Datasets}} 
We evaluate our proposed method on several benchmark datasets for node classification and link prediction tasks. For node classification, we use Citeseer, Pubmed, Corafull, Amazon-Photo, and OGBN-Proteins. For link prediction, we use Cora, Citeseer, Pubmed, ogbl-collab, and ogbl-ppa. The statistics of these datasets are summarized in Table~\ref{tab:dataset_statistics}. Each dataset is randomly split into training, validation, and test sets, following the standard protocols used in previous works \cite{wu2022nodeformer, shomer2024lpformer}.

\begin{table}[h]
  \caption{Statistics of benchmark datasets.}
  \label{tab:dataset_statistics}
  \resizebox{\linewidth}{!}{
  \begin{tabular}{c|lcccc}
    \toprule
    Task &Dataset & \#Nodes & \#Edges & \#Features & \#Classes \\
    \midrule
    \multirow{3}{*}{\makecell{NC\\+\\LP}} & Cora & 2,708 & 5,429 & 1,433 & 7 \\
    \multicolumn{1}{c|}{} & Citeseer & 3,327 & 4,732 & 3,703 & 6 \\
    \multicolumn{1}{c|}{} & Pubmed & 19,717 & 44,324 & 500 & 3 \\
    \midrule
    \multirow{3}*{NC} & Corafull & 19,793 & 65,311 & 8,710 & 70 \\
    \multicolumn{1}{c|}{} & Amazon-Photo & 7,650 & 119,081 & 745 & 8 \\
    \multicolumn{1}{c|}{} & OGBN-Proteins & 132,534 & 39,561,252 & 128 & 40 \\
    \midrule
    \multirow{2}*{LP} & OGBL-Collab & 235,868 & 1,285,465 & - & - \\
    \multicolumn{1}{c|}{} & OGBL-PPA & 576,289 & 30,326,273 & - & - \\
    \bottomrule
  \end{tabular}
  }
\end{table}

\begin{table*}[t]
  \caption{Performance comparison on node classification and link prediction tasks. \best{Red} highlights denote the best results, \second{blue} denotes the second best, and \third{green} denotes the third best. OOM indicates out-of-memory.}
  \label{tab:overall_performance}
  \resizebox{\linewidth}{!}{
  \begin{tabular}{>{\columncolor{rowsep}}c|c|ccccc|>{\columncolor{rowsep}}c|c|ccccc}
    \hline
    \multirow{14}{*}{} 
      & \multirow{3}{*}{Methods} & \multicolumn{5}{c|}{Datasets}
      & \multirow{14}{*}{} 
      & \multirow{3}{*}{Methods} & \multicolumn{5}{c}{Datasets} \\
    \cline{3-7}\cline{10-14}
    & & Citeseer & Pubmed & Corafull & Amazon-Photo & ogbn-proteins
      & & & Cora & Citeseer & Pubmed & ogbl-collab & ogbl-ppa \\
    \cline{3-7}\cline{10-14}
    & & ACC & ACC & ACC & ACC & ROC-AUC
      & & & MRR & MRR & MRR & Hits@50 & Hits@100 \\
    \cline{2-7}\cline{9-14}

    & GCN        & $76.50 \pm 1.36$ & $86.54 \pm 0.12$ & $61.76 \pm 0.14$ & $92.70 \pm 0.20$ & $72.51 \pm 0.35$
      & & GCN        & $32.50 \pm 6.87$ & $50.01 \pm 6.04$ & $19.94 \pm 4.24$ & $44.75 \pm 1.07$ & $18.67 \pm 1.32$ \\
    & GAT        & $76.55 \pm 1.23$ & $86.32 \pm 0.16$ & $64.47 \pm 0.18$ & $93.87 \pm 0.11$ & $72.02 \pm 0.44$
      & & GraphSAGE  & \third{$37.83 \pm 7.75$} & $47.84 \pm 6.39$ & $22.74 \pm 5.47$ & $48.10 \pm 0.81$ & $16.55 \pm 2.40$ \\
    & APPNP      & $76.53 \pm 1.16$ & $88.43 \pm 0.15$ & $65.16 \pm 0.28$ & $94.32 \pm 0.14$ & OOM
      & & GAE        & $29.98 \pm 3.21$ & $63.33 \pm 3.14$ & $16.67 \pm 0.19$ & OOM & OOM \\
      \cline{9-14}
    & GPRGNN     & $77.13 \pm 1.67$ & $89.34 \pm 0.25$ & $67.12 \pm 0.31$ & $94.49 \pm 0.14$ & $75.68 \pm 0.49$
      & & SEAL       & $26.69 \pm 5.89$ & $39.36 \pm 4.99$ & $38.06 \pm 5.18$ & $64.74 \pm 0.43$ & $48.80 \pm 3.16$ \\
    & PPRGo      & $76.23 \pm 1.02$ & $87.38 \pm 0.11$ & $63.54 \pm 0.25$ & $93.61 \pm 0.12$ & OOM
      & & NBFNet     & $37.69 \pm 3.97$ & $38.17 \pm 3.06$ & \best{$44.73 \pm 2.12$} & OOM & OOM \\
    \cline{2-7}

    & GraphGPS   & $76.99 \pm 1.12$ & $88.94 \pm 0.16$ & $55.76 \pm 0.23$ & $95.06 \pm 0.13$ & $76.83 \pm 0.26$
      & & Neo-GNN    & $22.65 \pm 2.60$ & $53.97 \pm 5.88$ & $31.45 \pm 3.17$ & $57.52 \pm 0.37$ & $49.13 \pm 0.60$ \\
    & NodeFormer & $76.33 \pm 0.59$ & $89.32 \pm 0.25$ & \second{$72.25 \pm 0.31$} & $93.46 \pm 0.35$ & \third{$77.45 \pm 1.15$}
      & & BUDDY      & $26.40 \pm 4.40$ & $59.48 \pm 8.96$ & $23.98 \pm 1.55$ & $65.94 \pm 0.58$ & $49.85 \pm 0.20$ \\
    & NAGphormer & \third{$77.42 \pm 1.41$} & \third{$89.70 \pm 0.19$} & $71.51 \pm 0.13$ & \third{$95.49 \pm 0.11$} & $73.61 \pm 0.33$
      & & NCN        & $32.93 \pm 3.80$ & $54.97 \pm 6.35$ & $35.65 \pm 4.60$ & $64.76 \pm 0.87$ & $61.19 \pm 0.85$ \\
    & Exphormer  & $76.83 \pm 1.24$ & $89.52 \pm 0.54$ & $69.09 \pm 0.72$ & $95.27 \pm 0.42$ & $74.58 \pm 0.26$
      & & NCNC       & $29.01 \pm 3.83$ & \third{$64.03 \pm 3.67$} & $25.70 \pm 4.82$ & \third{$66.61 \pm 0.71$} & \third{$61.42 \pm 0.73$} \\
    \cline{9-14}
    & GQT        & \best{$77.84 \pm 0.94$} & \second{$90.14 \pm 0.16$} & \third{$71.81 \pm 0.21$} & \second{$95.73 \pm 0.18$} & \best{$82.13 \pm 0.34$}
      & & LPFormer   & \second{$39.42 \pm 5.78$} & \second{$65.42 \pm 4.65$} & \third{$40.17 \pm 1.92$} & \best{$68.14 \pm 0.51$} & \best{$63.32 \pm 0.63$} \\
    \cline{2-7}\cline{9-14}
    \multirow{-14}{*}{\rotatebox[origin=c]{90}{\textbf{Task: Node Classification}}}
      & TAU (ours) & \second{$77.43 \pm 0.86$} & \best{$90.18 \pm 0.75$} & \best{$85.62 \pm 0.67$} & \best{$96.03 \pm 0.48$} & \second{$80.12 \pm 0.29$}
      & \multirow{-14}{*}{\rotatebox[origin=c]{90}{\textbf{Task: Link Prediction}}}
      & TAU (ours) & \best{$44.86 \pm 3.62$} & \best{$65.49 \pm 3.47$} & \second{$41.92 \pm 2.18$} & \second{$67.67 \pm 0.41$} & \second{$62.86 \pm 0.80$} \\
    \hline
  \end{tabular}
  }
\end{table*}

\subsubsection*{\textbf{Tasks}}
To comprehensively assess the proposed tokenizer's representation capability, we evaluate it on two fundamental graph tasks: node classification and link prediction.
\begin{itemize}[leftmargin=*]
  \item \textbf{Node Classification (NC)} focuses on evaluating how well the tokenizer preserves discriminative node information that reflects both local features and higher-order structural patterns. In this task, each node serves as an instance whose categorical label is inferred from its encoded representation. A tokenizer with strong expressiveness should enable downstream classifiers to separate nodes across different categories based on their quantized embeddings, reflecting its ability to retain meaningful node-level semantics after tokenization.
  \item \textbf{Link Prediction (LP)} examines the tokenizer's capacity to encode relational dependencies among nodes. Instead of classifying individual nodes, this task evaluates whether the learned tokens capture latent connectivity patterns that indicate the likelihood of edges existing between node pairs. Successful link prediction implies that the tokenizer effectively embeds structural proximity and interaction cues, enabling the reconstruction of missing or unobserved links from tokenized representations.
\end{itemize}

\subsubsection*{\textbf{Metrics}}
For node classification, we use accuracy (ACC) for datasets with balanced classes (CiteSeer, PubMed, CoraFull, Amazon-Photo) and ROC-AUC for the OGBN-Proteins dataset, which has imbalanced classes. For link prediction, Mean Reciprocal Rank (MRR) and Hits@K (K=50, 100) are used to evaluate model performance. 
We report the average and standard deviation of each metric over 10 independent runs with different random seeds to ensure the robustness of our results.

\subsubsection*{\textbf{Baselines}}
For the \textit{node classification} task, we compare our method with five representative graph neural network (GNN) baselines: GCN \cite{kipf2016gcn}, GAT \cite{velickovic2017gat}, APPNP \cite{gasteiger2018predict}, GPRGNN \cite{chien2020adaptive}, and PPRGo \cite{bojchevski2020scaling}. We also include five recent graph transformer-based GFMs: GraphGPS \cite{rampavsek2022recipe}, NodeFormer \cite{wu2022nodeformer}, NAGphormer \cite{chen2022nagphormer}, Exphormer \cite{shirzad2023exphormer}, and GQT \cite{wang2025learning}. For the \textit{link prediction} task, we compare against a broad set of baselines that cover both heuristic and learning-based approaches. As heuristic baselines, we use embedding-based GNN encoders with simple pairwise decoders, including GCN \cite{kipf2016gcn}, GraphSAGE \cite{hamilton2017inductive}, and GAE \cite{kipf2016gae}. We further include structure-centric methods that rely on local subgraphs or efficient structural features, including SEAL \cite{zhang2018link}, NBFNet \cite{zhu2021neural}, Neo-GNN \cite{yun2021neo}, BUDDY \cite{chamberlain2022graph}, NCN \cite{wang2023neural}, and NCNC \cite{wang2023neural}. Finally, we evaluate a recent graph transformer-based GFM, LPFormer \cite{shomer2024lpformer}, designed for link prediction. Detailed descriptions of all baselines are provided in Appendix \ref{appendix:baselines}.

\subsection{Overall Performance (RQ1)}
We evaluate the proposed method on widely used node classification and link prediction benchmarks. The comparison covers classical GNNs, transformer-style graph models, and recent quantized/tokenizer-based methods. In contrast to tokenizers that use a fixed hierarchy, TAU exposes multi-scale tokens. Then it applies a task-dependent gating on the tokenizer side to reweight different hierarchy levels before forming the final embedding. This design aligns with our goal: maintain a unified discrete token interface for pre-training and inference while adapting the effective structural scale to the downstream objective without changing the backbone encoder. Table~\ref{tab:overall_performance} reports the main results, and we discuss the key findings below.

\subsubsection*{\textbf{Node Classification}}
Across five node classification datasets, our approach consistently achieves first-tier performance and achieves the best result on three datasets (PubMed, Cora-Full, and Amazon-Photo). The most significant gain appears on Corafull, where TAU reaches $85.62\%$ ACC, while the strongest baseline in the table remains far lower. This gap suggests that a fixed tokenization hierarchy or a fixed structural bias is insufficient for graphs with richer, more diverse local-to-global patterns; TAU benefits from exposing hierarchical tokens and letting the task-specific gate decide how much each level should contribute. On PubMed and Amazon Photo, TAU also ranks at the top, achieving 90.18\% and 96.03\% ACC, respectively, with close competition from strong transformer-style and quantized baselines. On Citeseer, TAU does not take the single best score, but it still achieves first-tier performance: $77.43\%$ ACC, within $0.41$ points of the best method. This pattern aligns with our intent: the adaptive use of hierarchy yields robust results across datasets rather than overfitting to a single fixed granularity setting. On the large-scale OGBN-Proteins benchmark, TAU achieves $80.12\%$ ROC-AUC, which is slightly below the best score in the table but remains clearly competitive and in the first tier. Importantly, several baselines in the table run into OOM on this dataset, while TAU remains feasible. This supports our claim that a quantized discrete interface, paired with task-adaptive hierarchical weighting, can balance accuracy and practical resource cost on large graphs.

\subsubsection*{\textbf{Link Prediction}}
In link prediction, TAU again shows consistent first-tier performance across all datasets and achieves the best results on two out of five benchmarks (Cora and Citeseer). On Cora, TAU reaches $44.86\%$ MRR, improving over the strongest baseline in the table. On Citeseer, TAU achieves $65.49\%$ MRR, slightly higher than the best competing method. These improvements indicate that learning discrete tokens is not only helpful for node-level objectives but can also support edge-level reasoning when the tokenizer is allowed to adapt its effective hierarchy to the task. On the large-scale ogbl-collab and ogbl-ppa benchmarks, TAU achieves $67.67\%$ Hits@50 and $62.86\%$ Hits@100, respectively, which are both second best and within $0.5$ points of the best method. Notably, multiple baselines in the table encounter OOM on these large datasets, whereas TAU remains runnable. This reinforces the practicality of our tokenization design for large graphs.

A key reason that TAU transfers well from node classification to link prediction is that its adaptation happens before the final embedding is formed: the hierarchical codebook outputs multi-scale discrete tokens, and the gating reweights these levels based on the task. In addition, TAU incorporates edge-level training signals and relational positional encoding (RPE) to encode node-to-link relative structure, which is directly aligned with link prediction objectives. Overall, the results show that TAU achieves strong accuracy on both node-level and edge-level tasks under a unified discrete token interface and remains stable in large-scale settings where several baselines fail due to memory limits.

\subsection{Ablation Study (RQ2)}
\begin{figure}[t]
  \centering
  \begin{minipage}{\linewidth}
    \centering
    \includegraphics[width=\linewidth]{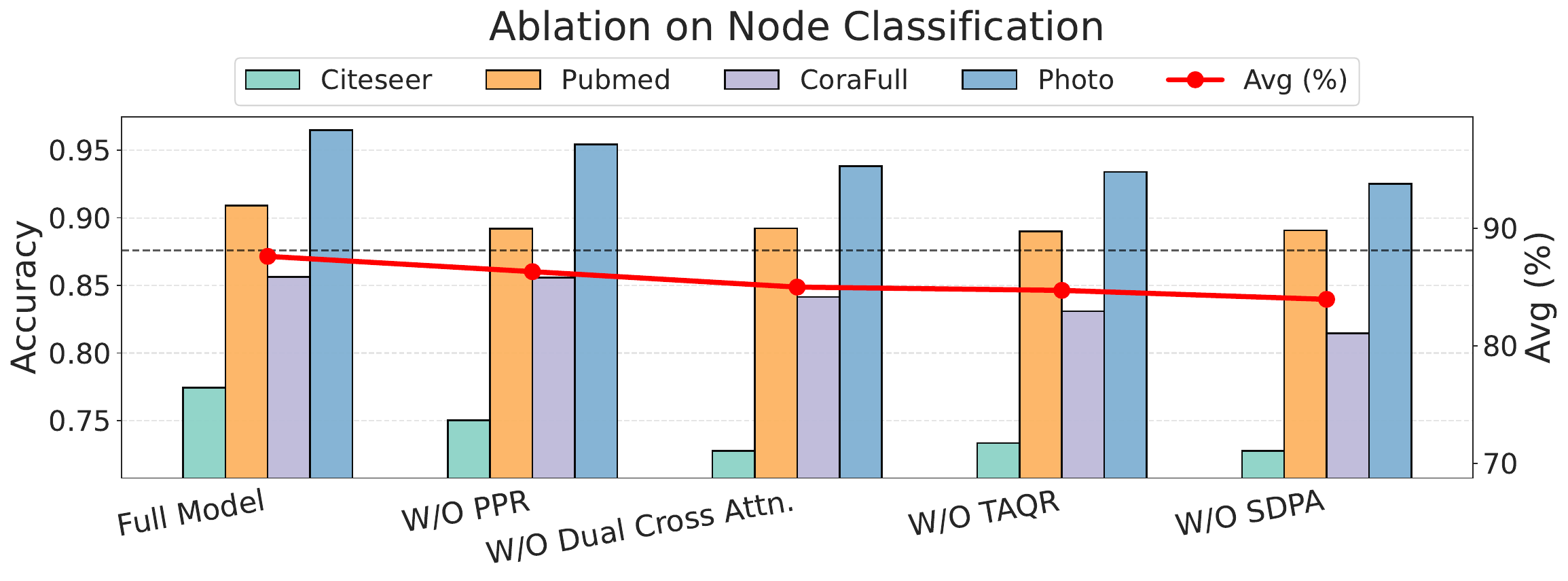} \\
    \includegraphics[width=\linewidth]{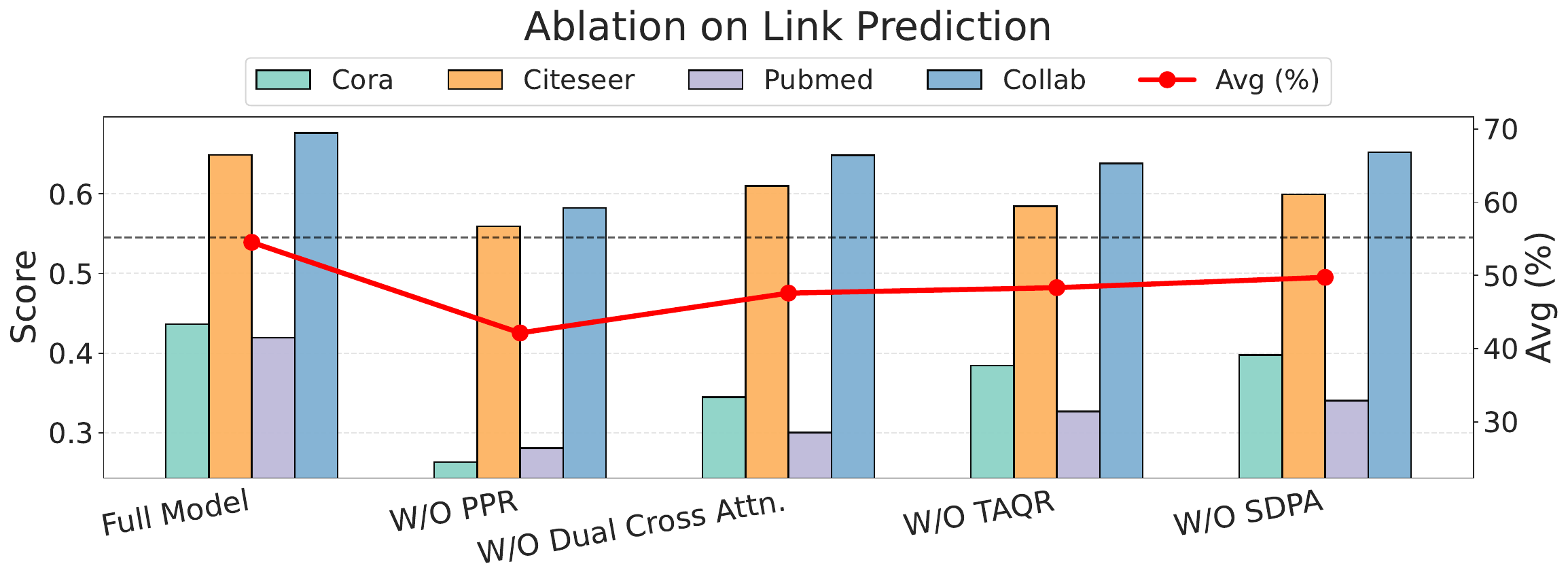}
  \end{minipage}
  \caption{Ablation study of different components in our proposed TAU for node classification and link prediction.}
  \label{fig:ablation}
\end{figure}

To investigate the impact of each proposed component in TAU on its performance, we develop the following model variants for an ablation study:
\begin{itemize}[leftmargin=*]
  \item \textbf{(-) PPR.} This variant removes the PPR branch and relies only on the remaining structural and token streams for prediction.
  \item \textbf{(-) Dual Cross Attn.} This variant removes the dual cross-attention fusion module and combines local and global information without explicit cross-stream interaction.
  \item \textbf{(-) TAQR.} This variant removes task-adaptive quantization routing and trains the tokenizer without task-guided routing and its associated regularization effects.
  \item \textbf{(-) SPDA.} This variant replaces head-specific gated sparse attention with standard multi-head self-attention, removing head-wise gating and sparsification.
\end{itemize}
We report the ablation results of TAU and its variants on both node classification (NC) and link prediction (LP) in Figure~\ref{fig:ablation}. In summary, our findings are as follows:
\begin{enumerate}[leftmargin=*]
  \item The PPR branch is most critical for LP, while it has a smaller and less consistent effect on NC. LP depends more on multi-hop structural evidence because edge reasoning often requires linking two nodes through longer-range paths and shared neighborhoods. In contrast, NC can often be solved with more local cues, so removing explicit multi-hop signals tends to cause less harm, especially when label information is concentrated in short-range neighborhoods.
  \item Dual cross-attention fusion improves both tasks by stabilizing NC and improving semantic alignment for LP. The broad drop after removing it suggests that having both local and global streams is not sufficient on its own. Cross-attention provides a controlled way for the two streams to exchange information, reducing stream imbalance and mismatch between the hierarchy levels each task requires.
  \item TAQR supports performance in both NC and LP by keeping discrete tokens informative and diverse. Without task-adaptive routing, the tokenizer has a weaker ability to assign capacity to the hierarchy levels that matter for the current objective. This reduces the codebook's usefulness across tasks, suggesting that task-guided token usage is important for a shared tokenizer.
  \item SPDA contributes beyond efficiency, with a clear benefit for NC and a consistent benefit for LP. Replacing SPDA with standard multi-head attention increases redundancy in token interactions and weakens suppression of less relevant hierarchy levels. This indicates that head-specific gating and sparsification help the model focus on task-relevant structural ranges, which aligns with our goal of adapting hierarchy usage within a unified tokenization framework.
\end{enumerate}

\subsection{Task-Adaptive Routing Weight Analysis (RQ3)}
\begin{figure}[t]
  \centering
  \begin{minipage}{\linewidth}
    \centering
    \includegraphics[width=0.49\linewidth]{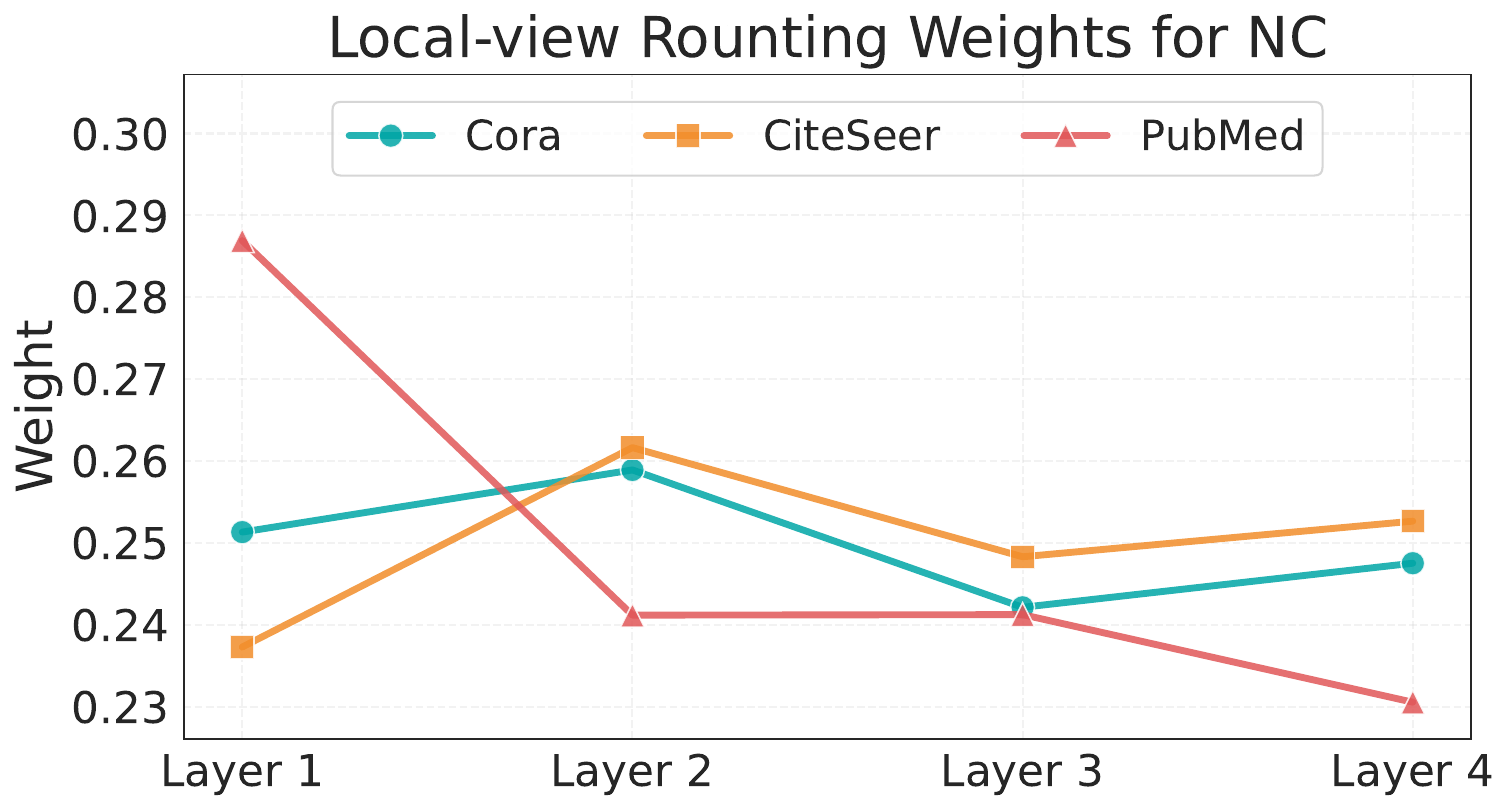}
    \includegraphics[width=0.49\linewidth]{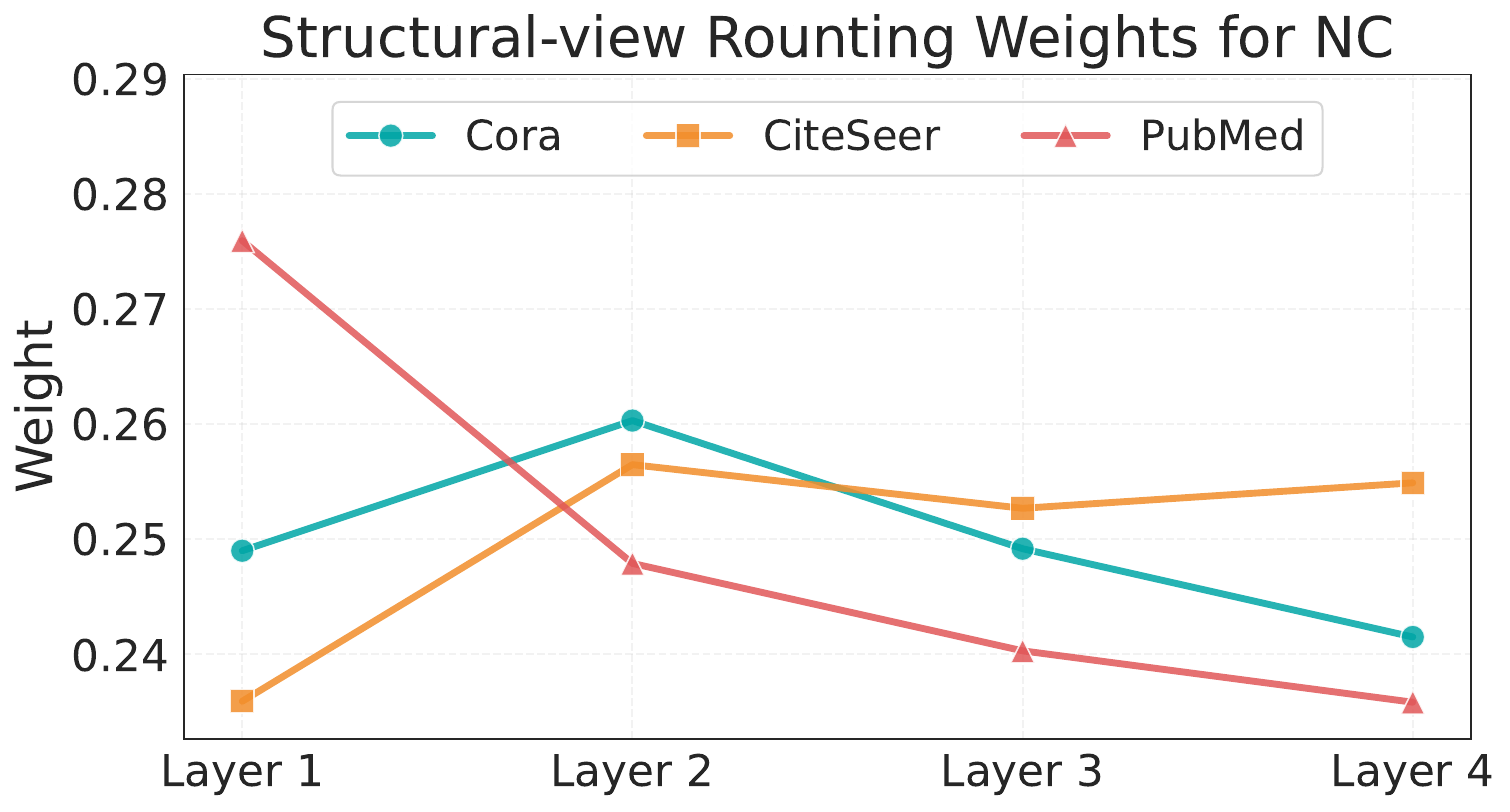}\\
    \includegraphics[width=0.49\linewidth]{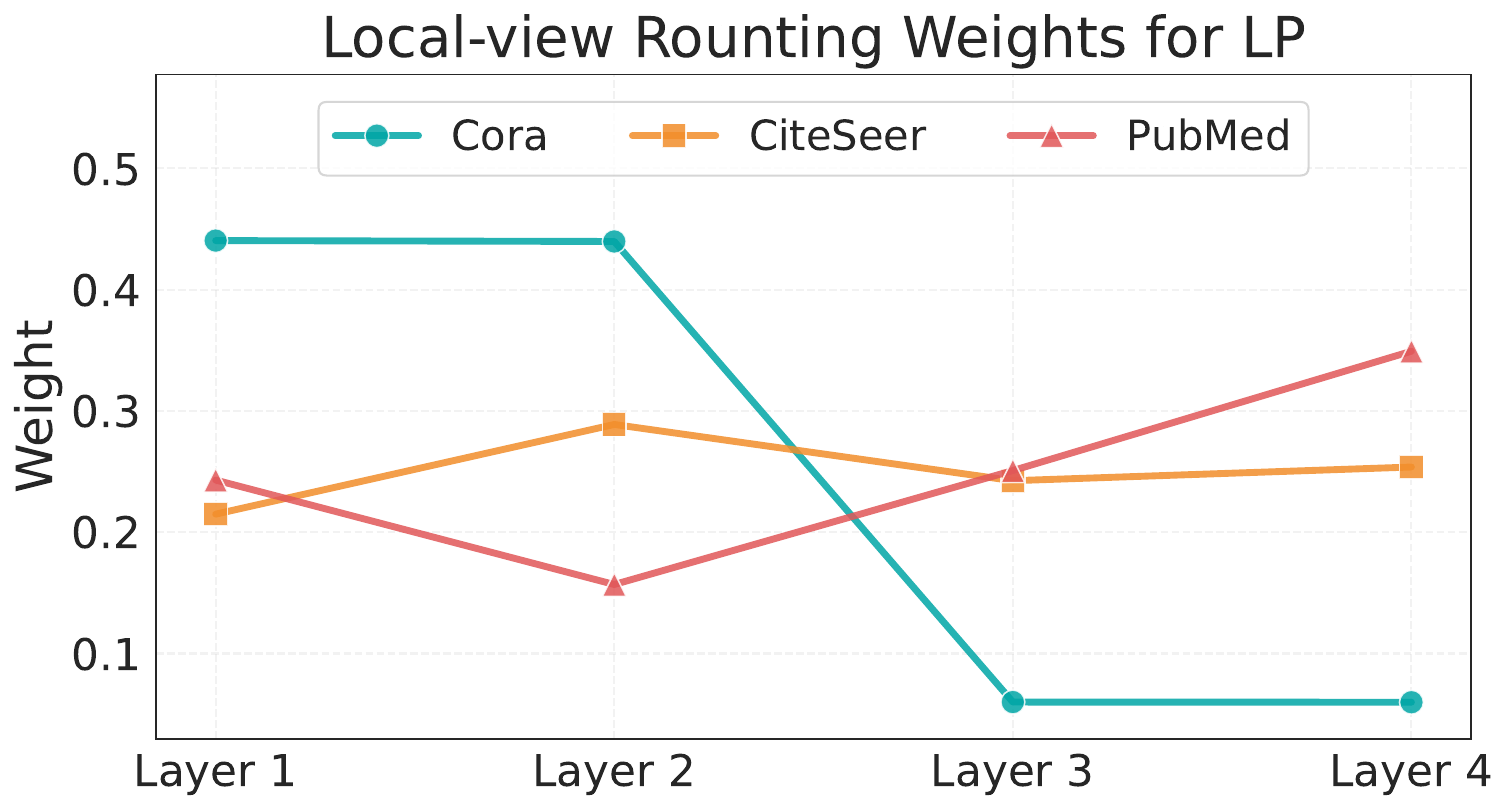}
    \includegraphics[width=0.49\linewidth]{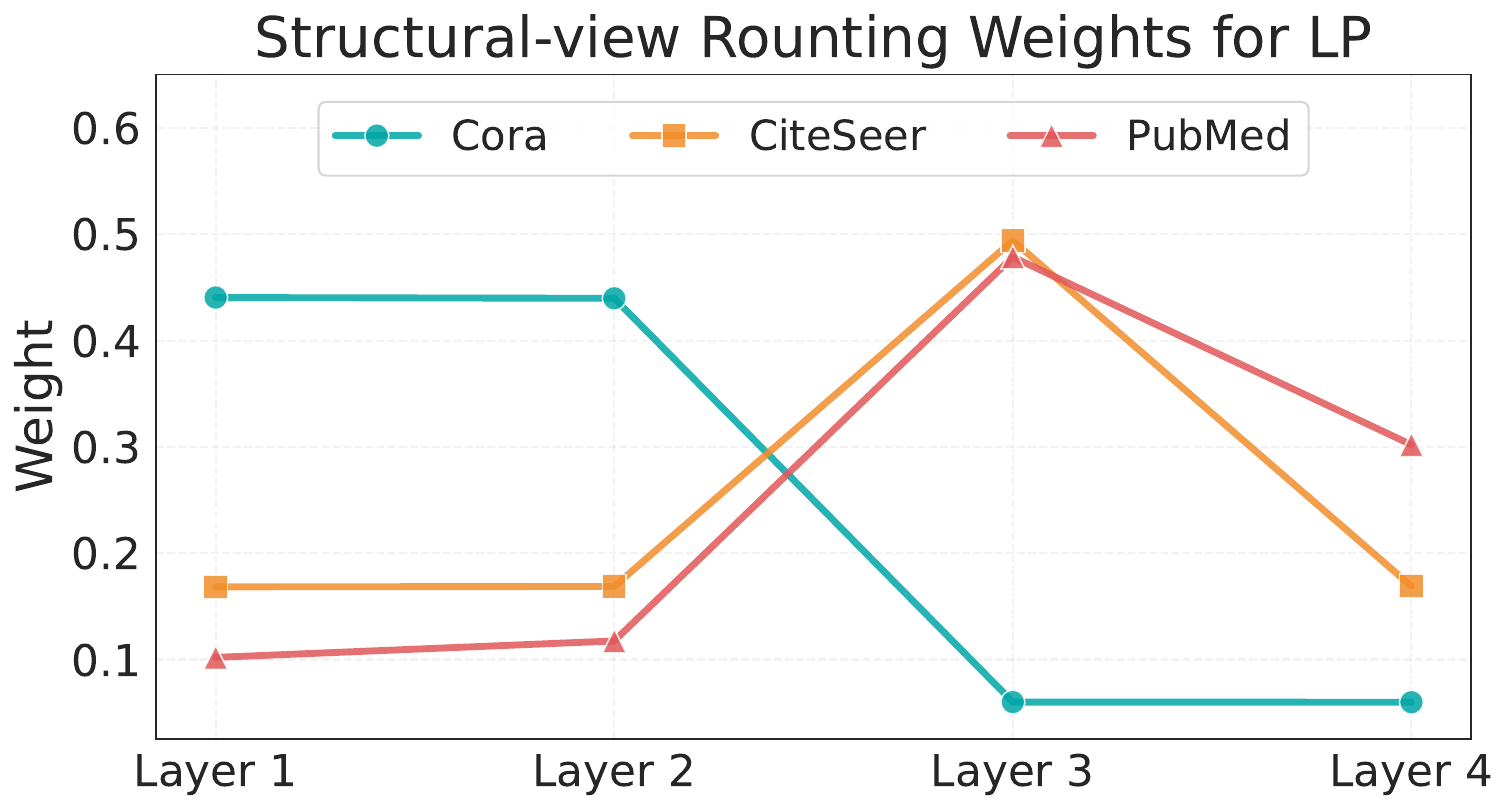}
  \end{minipage}
    \caption{Analysis of task-adaptive routing weights learned by TAQR on different datasets for node classification and link prediction.}
    \label{fig:routing_weight}
\end{figure}

To study how TAQR uses the hierarchy across tasks and datasets, we analyze the learned routing weights at each hierarchy level after training. Figure~\ref{fig:routing_weight} reports the averaged level-wise routing weights for node classification (top) and link prediction (bottom) on three datasets, using two token streams: a local semantic stream from raw node representations and a multi-hop structural stream from PPR-propagated representations. For node classification, routing weights are near-uniform across all four levels on all datasets for both streams, suggesting that node supervision benefits from combining multiple granularities rather than relying on a single level. For link prediction, the model shows clearer, dataset-dependent preferences that also vary by stream. Some datasets emphasize shallow levels, consistent with strong short-range and local semantic cues, while others assign more weight to mid or deeper levels, especially in the structural stream, consistent with a stronger need for longer-range connectivity. The semantic stream often stays more distributed even when the structural stream peaks, indicating that TAQR can set the effective granularity separately for semantic and structural information, though in some cases the two streams align. Overall, these results show that TAQR adapts hierarchy usage to the task and graph, rather than acting as a fixed hierarchy mixer. This supports our claim that a unified discrete token interface should not assume a single granularity and helps explain why TAU remains robust across node- and link-level benchmarks under the same tokenizer design.\looseness=-1

\subsection{Task-Adaptability Analysis (RQ4)}
\begin{figure}[t]
  \centering
  \begin{minipage}{\linewidth}
    \centering
    \includegraphics[width=0.7\linewidth]{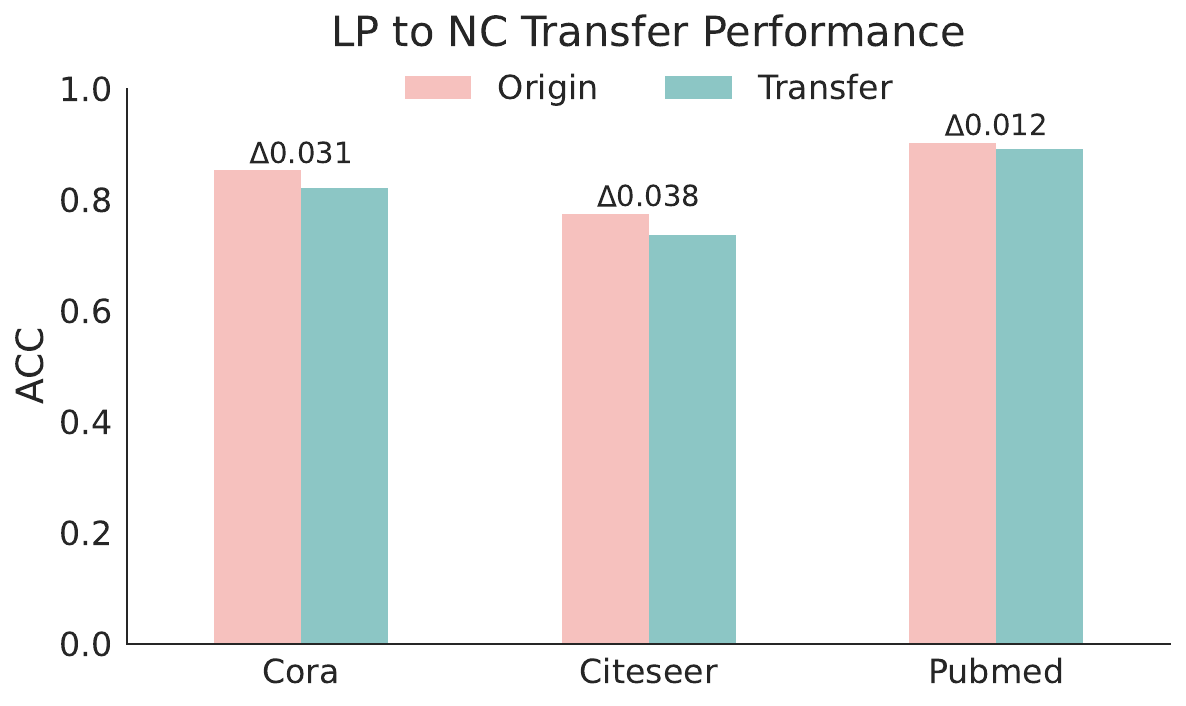}\\
    \includegraphics[width=0.7\linewidth]{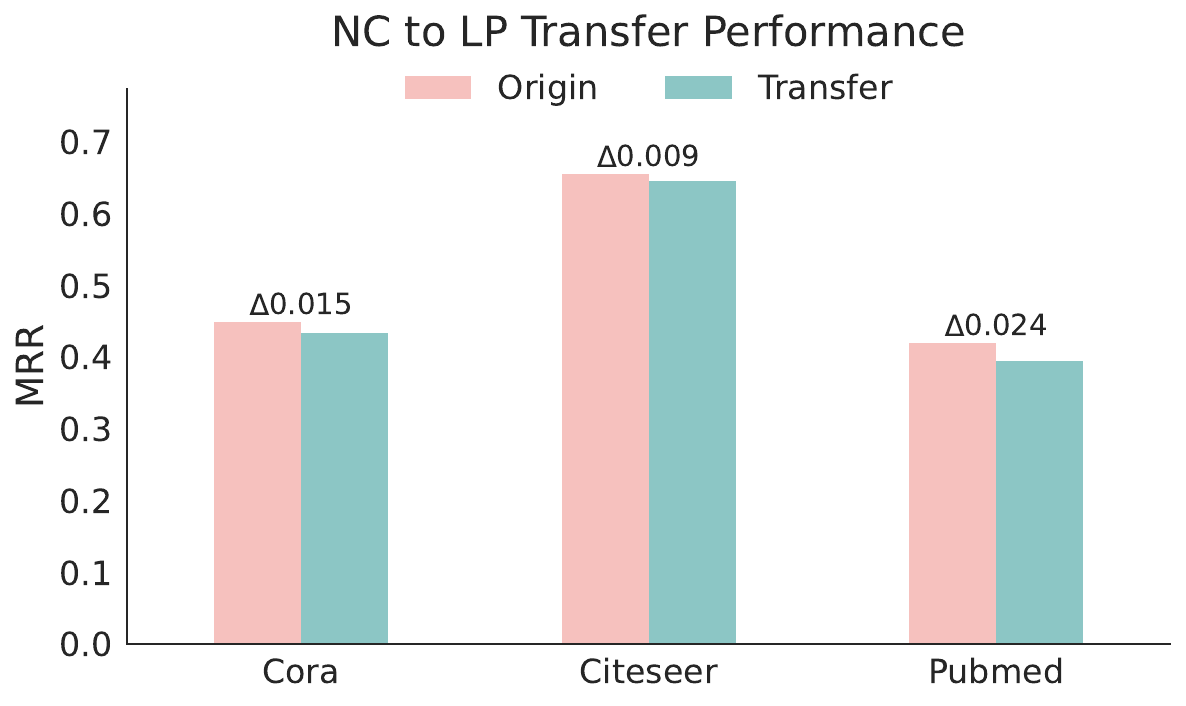}
  \end{minipage}
    \caption{Task-adaptability analysis of our proposed TAU tokenizer through transfer learning between node classification and link prediction tasks.}
    \label{fig:transfer}
  \vspace{-2em}
\end{figure}

To assess TAU's task adaptability, we conduct a transfer study between node classification (NC) and link prediction (LP). We first pre-train TAU on one task, then freeze all tokenizer parameters, and fine-tune only the dual cross-attention fuse gate, the Graph Transformer backbone, and task-specific adapters on the target task. We report the \emph{relative decrease rate} after transfer, measured against training-from-scratch on the target task. Figure~\ref{fig:transfer} shows the trends on three datasets (Cora, Citeseer, Pubmed). When transferring from LP to NC, the decrease rates are small: $3.64\%$ (Cora), $4.96\%$ (Citeseer), and $1.30\%$ (Pubmed). The average decrease rate is $3.30\%$, with a worst-case decrease of $4.96\%$. This indicates that the frozen discrete tokenizer trained on LP still produces graph tokens that remain useful for node-level supervision, and that target-task adaptation can be handled largely by the lightweight gate/backbone/adapters. When transferring from NC to LP, the decrease rates remain low but show a slightly larger spread: $3.34\%$ (Cora), $1.36\%$ (Citeseer), and $5.68\%$ (Pubmed). The average decrease rate is $3.46\%$, and the worst-case decrease is $5.68\%$. This asymmetry suggests that LP may require more task-specific relational cues (e.g., edge-oriented structure signals) than NC, so a tokenizer pre-trained on NC may miss some LP-focused information. Still, all decreases remain within about $6\%$, which supports our core claim: TAU learns largely task-shared hierarchical tokens, while task differences are mainly absorbed by the adaptive fusion gate and the downstream modules, rather than retraining or changing the tokenizer.

\subsection{Efficiency Analysis (RQ5)}
In this section, we study the training efficiency of the task-adaptive routing module in TAU, Task-Adaptive Quantization Routing. Our goal here is to verify whether TAQR improves optimization and reduces the total training cost. To this end, we compare the full TAU model (with TAQR) against an ablated variant that removes TAQR while keeping the tokenizer, codebooks, and backbone unchanged. We report training time, peak memory usage, and convergence behavior on multiple datasets for both node classification and link prediction, with detailed numbers summarized in Table~\ref{tab:efficiency}. Across settings, TAQR improves convergence speed and reduces the number of epochs needed to reach the best checkpoint. For example, in PubMed node classification, the full model converges in 277 epochs, whereas the variant without TAQR requires 293 epochs to achieve a similar accuracy. This behavior is consistent with TAQR's role in our design: TAQR reweights and routes information across different quantization levels based on the task signal, so the model can rely on earlier levels of the hierarchy that are more useful for the current objective. As a result, optimization becomes easier, and the model spends fewer epochs exploring less informative token combinations. TAQR incurs a small overhead per epoch due to an extra routing step on the tokenization side. However, Table~\ref{tab:efficiency} shows that the reduction in epochs typically leads to lower total wall-clock time, so the net training cost decreases. We also observe that TAQR stabilizes early training, leading to smoother loss curves and fewer fluctuations in validation scores. We attribute this to more stable token usage across hierarchy levels: instead of forcing the backbone to handle a fixed mixture of multi-scale tokens, TAQR adjusts the effective contribution of each level, which reduces gradient noise from mismatched or redundant hierarchy signals. These results indicate that TAQR is not only an accuracy module but also an efficiency module. It improves hierarchical quantization utilization through task-aware routing, accelerating convergence and stabilizing training across both node-level and edge-level tasks, while keeping the GFM backbone unchanged.

\begin{table}[t]
  \caption{Efficiency comparison between the full model (with TAQR) and the variant without TAQR. Training time is the wall-clock time to reach the best checkpoint used for reporting performance. Peak GRAM is the maximum GPU RAM usage during training.}
  \label{tab:efficiency}
  \resizebox{\linewidth}{!}{
  \begin{tabular}{c|c|c|ccc}
    \hline
    Task & Datasets & Components & Epochs & Training Time & Peak GRAM\\
    \hline
    \multirow{6}*{NC} & \multirow{2}*{Pubmed} & Full Model & 277 & 00h 00m 16s & 1.75 GB \\
    \multicolumn{1}{c|}{} & \multicolumn{1}{c|}{} & w/o TAQR & 293 & 00h 00m 19s & 1.75 GB \\
    \cline{2-6}
    \multicolumn{1}{c|}{} & \multirow{2}*{Amazon Photo} & Full Model & 163 & 00h 00m 23s & 4.36 GB \\
    \multicolumn{1}{c|}{} & \multicolumn{1}{c|}{} & w/o TAQR & 178 & 00h 00m 26s & 4.35 GB \\
    \cline{2-6}
    \multicolumn{1}{c|}{} & \multirow{2}*{ogbn-proteins} & Full Model & 387 & 00h 20m 11s & 9.76 GB \\
    \multicolumn{1}{c|}{} & \multicolumn{1}{c|}{} & w/o TAQR & 500 & 00h 26m 05s & 9.74 GB \\
    \hline
    \multirow{6}*{LP} & \multirow{2}*{Pubmed} & Full Model & 93 & 00h 02m 42s & 2.94 GB \\
    \multicolumn{1}{c|}{} & \multicolumn{1}{c|}{} & w/o TAQR & 150 & 00h 03m 51s & 2.93 GB \\
    \cline{2-6}
    \multicolumn{1}{c|}{} & \multirow{2}*{ogbl-collab} & Full Model & 66 & 00h 07m 21s & 14.15 GB \\
    \multicolumn{1}{c|}{} & \multicolumn{1}{c|}{} & w/o TAQR & 85 & 00h 09m 17s & 14.12 GB \\
    \cline{2-6}
    \multicolumn{1}{c|}{} & \multirow{2}*{ogbl-ppa} & Full Model & 75 & 10h 02m 44s & 29.88 GB \\
    \multicolumn{1}{c|}{} & \multicolumn{1}{c|}{} & w/o TAQR & 102 & 13h 08m 21s & 29.67 GB \\
    \hline
  \end{tabular}
  }
\end{table}

\section{Conclusion \& Future Work}
In this paper, we propose TAU, a quantized hierarchical self-weighted tokenizer for graph foundation models. TAU generates multi-scale discrete graph tokens through a dual-branch quantization mechanism that captures both local and global structural information. A task-adaptive quantization routing module is introduced to dynamically reweight the contributions of different hierarchy levels based on downstream tasks. Extensive experiments on node classification and link prediction benchmarks demonstrate that TAU achieves strong performance across tasks while maintaining efficiency and scalability. Ablation studies confirm the effectiveness of each component in our design, and transfer learning analyses highlight the learned tokenizer's task adaptability. Overall, TAU provides a unified, flexible tokenization framework that enhances the versatility of graph foundation models across diverse applications. Future work includes exploring more sophisticated routing mechanisms, such as dynamic routing or reinforcement learning-based approaches, to further improve the adaptability of the tokenizer. We also plan to investigate the application of TAU in other graph tasks, such as graph classification and community detection, to evaluate its generality.\looseness=-1

\appendix
\section{Task Coverage of Recent Graph Foundation Models}
\label{appendix:task_coverage}
Table \ref{tab:gfm_task_coverage} summarizes the task coverage of recent GFMs and representative graph Transformers. Most existing GFMs focus on either node classification or link prediction, with limited support for both tasks within a single model. Our proposed TAU tokenizer addresses this gap by providing a unified tokenization framework that can adaptively handle multiple downstream tasks, enhancing the versatility and applicability of graph foundation models.
\begin{table}[h]
  \centering
  \small
  \caption{Task coverage of recent GFMs and representative graph Transformers. \cmark: full experiments reported for the task; \xmark: not supported or not discussed; \pmark: the paper does not conduct comprehensive experiments. NC: node classification; LP: link prediction; GC: graph classification; GR: graph regression.}
  \label{tab:gfm_task_coverage}
  \resizebox{\linewidth}{!}{
    \begin{tabular}{lccccc}
      \toprule
      \textbf{Method} & \textbf{Venue} & \textbf{NC} & \textbf{LP} & \textbf{GC} & \textbf{GR} \\
      \midrule
      GraphGPS \cite{rampavsek2022recipe}     & NeurIPS 2022  & \cmark & \pmark & \cmark & \cmark \\
      NodeFormer \cite{wu2022nodeformer}      & NeurIPS 2022  & \cmark & \xmark & \xmark & \xmark \\
      Exphormer \cite{shirzad2023exphormer}   & ICML 2023     &\cmark & \pmark & \xmark & \xmark \\
      NAGphormer \cite{chen2022nagphormer}    & ICLR 2023     & \cmark & \xmark & \xmark & \xmark \\
      LPFormer \cite{shomer2024lpformer}      & SIGKDD 2024   & \xmark & \cmark & \xmark & \xmark \\
      GQT \cite{wang2025learning}             & ICLR 2025     & \cmark & \pmark & \xmark & \xmark \\
      \bottomrule
    \end{tabular}
  }
\end{table}

\section{Performance on Graph-level Tasks}
\label{appendix:graph_level_tasks}
While our main experiments focus on node classification and link prediction, we further evaluate TAU on graph-level tasks to test its broader applicability. We conduct experiments using GRIT \cite{ma2023graph} with TAU as the tokenizer, and report results on three benchmarks following the setup in \cite{dwivedi2023benchmarking}. We also include GraphGPS \cite{rampavsek2022recipe} and EGT \cite{hussain2022global}, two recent graph transformer-based graph foundation model, as baselines. The results are summarized in Table \ref{tab:graph_level_performance}. When TAU is used as the tokenizer, GRIT achieves better performance, showing that TAU is also effective for graph-level tasks. In particular, TAU-GRIT outperforms EGT, GraphGPS and GRIT across all three benchmarks, suggesting that TAU's adaptive hierarchical tokenization improves graph-level representation learning.

\begin{table}[h]
  \centering
  \small
  \caption{Performance of TAU on graph-level tasks. We report the average and standard deviation over 5 runs.}
  \label{tab:graph_level_performance}
  \resizebox{\linewidth}{!}{
    \begin{tabular}{lccc}
      \toprule
      \multirow{2}{*}{\textbf{Method}} & \textbf{ZINC} & \textbf{MNIST} & \textbf{CIFAR10} \\
       & MAE $\downarrow$ & ACC $\uparrow$ & ACC $\uparrow$ \\
      \midrule
      EGT & $0.108 \pm 0.009$ & $98.173 \pm 0.087$ & $68.702 \pm 0.409$ \\ 
      GraphGPS & $0.070 \pm 0.004$ & $98.051 \pm 0.126$ & $72.298 \pm 0.356$ \\
      GRIT & $0.059 \pm 0.002$ & $98.108 \pm 0.111$ & $76.468 \pm 0.881$ \\
      \hline
      TAU-GRIT & $\mathbf{0.051 \pm 0.001}$ & $\mathbf{98.213 \pm 0.098}$ & $\mathbf{77.784 \pm 0.490}$ \\
      \bottomrule
    \end{tabular}
  }
\end{table}

\section{Head-specific gated sparse attention: detailed formulation}
\label{appendix:gated_attention_details}
This appendix provides a complete formulation of head-specific gated sparse attention on a graph. The aggregation is a Scaled Dot-Product Attention (SDPA) computation restricted to graph edges, and the head-specific gate is applied multiplicatively after the SDPA output \cite{qiu2025gated}. For each target token $j$ and head $h$, attention is computed only over the incoming neighbor set $\mathcal{N}(j)=\{i:(i\rightarrow j)\in\mathcal{E}\}$.

Given the head query $q_j^h$, head key $k_i^h$, and head value $v_i^h$, we first compute biased scaled dot-product scores with temperature $\tau$:
\begin{equation}
s_{i j}^h=\frac{\left\langle q_j^h,\, k_i^h+b_{i j}\right\rangle}{\sqrt{d_k\,\tau}}.
\end{equation}
We then normalize these scores with a masked softmax over $\mathcal{N}(j)$:
\begin{equation}
\alpha_{i j}^h=\frac{\exp \left(s_{i j}^h\right)}{\sum_{i^{\prime}\in\mathcal{N}(j)} \exp \left(s_{i^{\prime} j}^h\right)}.
\end{equation}
The sparse SDPA output for head $h$ at token $j$ is the weighted sum of values:
\begin{equation}
\tilde{o}_j^h=\sum_{i\in\mathcal{N}(j)} \alpha_{i j}^h v_i^h.
\end{equation}

Following \cite{qiu2025gated}, we compute a head-specific sigmoid gate from the head query and apply it to the SDPA output:
\begin{equation}
g_j^h=\sigma\!\left(q_j^h\right),
\end{equation}
where $\sigma(\cdot)$ is applied elementwise. The gated head output is then
\begin{equation}
o_j^h=g_j^h \odot \tilde{o}_j^h.
\end{equation}

Finally, we concatenate head outputs and apply the output projection:
\begin{equation}
o_j=\operatorname{Concat}\left(o_j^1, \ldots, o_j^H\right)\in \mathbb{R}^D,
\end{equation}
\begin{equation}
a_j=W_o\, o_j.
\end{equation}
Here $b_{ij}$ is a structural bias term, $d_k$ is the per-head key/query dimension, $\tau$ is the temperature, $H$ is the number of heads, $D$ is the concatenated embedding dimension, and $W_o$ is the output projection matrix.

\section{Complexity Analysis}
\subsubsection*{\textit{\textbf{Space Complexity}}}
The space complexity of TAU is given by $\mathcal{O}(m + Nd + Nd_h + \sum_{m=1}^{M} k_m d_h + Hm)$, where $m$ is the number of edges, $N$ is the number of nodes, $d$ is the input feature dimension, $d_h$ is the embedding dimension, $M$ is the number of RVQ levels, $k_m$ is the number of codewords at level $m$, and $H$ is the number of attention heads. TAU stores the sparse adjacency structure and node features, requiring $\mathcal{O}(m)$ and $\mathcal{O}(Nd)$ space, respectively. It also stores intermediate continuous representations (the GCN embeddings $H$ and the PPR-enhanced features $H_{\mathrm{PPR}}$) with total storage $\mathcal{O}(Nd_h)$, and maintains the RVQ codebooks with storage $\mathcal{O}(\sum_{m=1}^{M} k_m d_h)$. For the backbone, the Graph Transformer uses sparse attention restricted to graph edges, so attention weights and related activations scale with the number of edges (up to $\mathcal{O}(Hm)$) rather than $\mathcal{O}(HN^2)$. In mini-batch training with neighbor sampling, TAU only materializes a sampled subgraph per batch, reducing peak activation storage from $m$ and $N$ to the sampled edge and node counts.

\subsubsection*{\textit{\textbf{Time Complexity}}}
The time complexity of TAU can be formulated as $\mathcal{O}(r((L+K)e_b d_h + Mb\bar{k}d_h + Hk^2 d_h + bk d_h + L_THe_b d_h))$, where $r$ is the number of epochs, $b$ is the mini-batch size, $e_b$ is the number of sampled edges, $L$ is the number of GCN layers, $K$ is the number of PPR propagation steps, $\bar{k}$ is a representative RVQ codebook size, $k$ is the fused codebook size for assignment, and $L_T$ is the number of Graph Transformer layers. The GCN encoder costs $\mathcal{O}(Le_b d_h)$ per batch due to sparse message passing, and PPR diffusion costs $\mathcal{O}(Ke_b d_h)$ because each step applies a sparse normalized adjacency operator. RVQ nearest-codeword selection costs $\mathcal{O}(Mb\bar{k}d_h)$ in a direct implementation. TAQR uses mean pooling, a two-layer MLP per depth, and a softmax over depths, which is lower order than RVQ in common settings. Dual cross-attention over the two token sets has cost $\mathcal{O}(Hk^2d_h)$. Codebook assignment compares each node with a fixed-size codebook, costing $\mathcal{O}(bk d_h)$ per batch and remaining independent of the full graph size beyond the processed nodes. Each Graph Transformer layer then applies sparse edge attention, giving $\mathcal{O}(L_THe_b d_h)$ per batch rather than $\mathcal{O}(L_THb^2 d_h)$.

\section{Experimental Configurations}
\label{appendix:exp_setup}
\subsection{Baselines Specifications}
\label{appendix:baselines}
\begin{itemize}[leftmargin=*]
    \item \textbf{GCN} \cite{kipf2016gcn} is a graph convolutional network that updates node features by neighborhood aggregation with a normalized adjacency operator;
    \item \textbf{GAT} \cite{velickovic2017gat} is a graph neural network that uses attention to assign different weights to different neighbors during aggregation;
    \item \textbf{APPNP} \cite{gasteiger2018predict} is a propagation scheme based on personalized PageRank that decouples prediction from propagation to better use multi-hop information;
    \item \textbf{GPRGNN} \cite{chien2020adaptive} is a generalized PageRank GNN that learns trainable propagation weights to combine information from multiple hops;
    \item \textbf{PPRGo} \cite{bojchevski2020scaling} is a scalable model that uses sparse approximations of personalized PageRank to precompute diffusion features for large graphs;
    \item \textbf{GraphGPS} \cite{rampavsek2022recipe} is a hybrid graph transformer that combines local message passing with global attention and structural or positional encodings in a modular design;
    \item \textbf{NodeFormer} \cite{wu2022nodeformer} is a scalable graph transformer that enables long-range interaction by learning latent graph structures with efficient attention-style computation;
    \item \textbf{NAGphormer} \cite{chen2022nagphormer} is a tokenized graph transformer that represents each node using a sequence built from multi-hop neighborhood aggregation so that training can be done in mini-batches;
    \item \textbf{Exphormer} \cite{shirzad2023exphormer} is a sparse graph transformer that uses structured sparse attention (for example, via expander edges and global nodes) to improve scaling while keeping accuracy competitive;
    \item \textbf{GQT} \cite{wang2025learning} is a graph quantized tokenizer approach that learns discrete graph tokens (for example, with residual vector quantization) to support transformer backbones with learned tokenization;
    \item \textbf{GCN} \cite{kipf2016gcn} is used as a message-passing encoder for link prediction, typically combined with a simple pairwise decoder (for example, inner product or an MLP);
    \item \textbf{GraphSAGE} \cite{hamilton2017inductive} is an inductive GNN that samples and aggregates neighborhood features to build node embeddings that generalize to unseen nodes;
    \item \textbf{GAE} \cite{kipf2016gae} is a graph auto-encoder that learns node embeddings with a GNN encoder and reconstructs edges with a simple decoder (often inner product);
    \item \textbf{SEAL} \cite{zhang2018link} is a subgraph-based link prediction method that extracts an enclosing subgraph around each candidate link and uses a GNN to classify link existence from local structure;
    \item \textbf{NBFNet} \cite{zhu2021neural} is a path-based link prediction framework that aggregates information over multiple paths between node pairs using a Bellman-Ford style dynamic program with learned operators;
    \item \textbf{Neo-GNN} \cite{yun2021neo} is a neighborhood-overlap-aware model that learns structural signals related to common neighbors and multi-hop overlap for link prediction;
    \item \textbf{BUDDY} \cite{chamberlain2022graph} is an efficient link prediction method that uses precomputed structural features (via subgraph sketching) to avoid explicit per-edge subgraph extraction;
    \item \textbf{NCN} \cite{wang2023neural} is a link predictor that uses learnable pairwise representations guided by structural features related to common neighbors;
    \item \textbf{NCNC} \cite{wang2023neural} extends NCN by compensating for missing edges in the observed graph so that common-neighbor-style signals are less biased by incompleteness;
    \item \textbf{LPFormer} \cite{shomer2024lpformer} is a link prediction model that uses attention to adaptively learn pairwise encodings for each candidate link, instead of using a fixed hand-designed encoding.
\end{itemize}

\subsection{Implementation Details}
We implement the proposed method using the PyTorch and PyTorch Geometric libraries. The GNN encoder in the tokenizer uses a GCN architecture, with the number of layers and hidden dimensions tunable hyperparameters. The number of quantization levels $M$ is fixed at 4, and the size of each codebook $k_m$ is determined by the dataset's scale. The temperature parameter $\tau$ in the self-weighted gate is set to 1.5. Both hyperparameters $\alpha$ and $\beta$ in the tokenizer loss are assigned a value of 0.01. For the graph foundation model, the number of transformer layers and attention heads is configured based on the dataset size and complexity. To reduce computational overhead during training, the PPR matrix is precomputed for each dataset before the first training run. We employ the Adam optimizer to optimize the model parameters. The models are trained for 500 epochs, with early stopping based on validation performance. All hyperparameter settings can be found in our codebase \url{https://anonymous.4open.science/r/TAU-0C1F}.

\subsection{Hardware and Software}
We conduct our experiments on a workstation equipped with an AMD Ryzen 9 9950X CPU, 64GB RAM, and a single NVIDIA RTX 5090 GPU with 32 GB memory. The software environment includes Python 3.13, PyTorch 2.9, and CUDA 13.0.

\newpage

\bibliographystyle{ACM-Reference-Format}
\balance
\bibliography{ref}


\begin{thebibliography}{76}


\ifx \showCODEN    \undefined \def \showCODEN     #1{\unskip}     \fi
\ifx \showISBNx    \undefined \def \showISBNx     #1{\unskip}     \fi
\ifx \showISBNxiii \undefined \def \showISBNxiii  #1{\unskip}     \fi
\ifx \showISSN     \undefined \def \showISSN      #1{\unskip}     \fi
\ifx \showLCCN     \undefined \def \showLCCN      #1{\unskip}     \fi
\ifx \shownote     \undefined \def \shownote      #1{#1}          \fi
\ifx \showarticletitle \undefined \def \showarticletitle #1{#1}   \fi
\ifx \showURL      \undefined \def \showURL       {\relax}        \fi
\providecommand\bibfield[2]{#2}
\providecommand\bibinfo[2]{#2}
\providecommand\natexlab[1]{#1}
\providecommand\showeprint[2][]{arXiv:#2}

\bibitem[Abu-El-Haija et~al\mbox{.}(2019)]%
        {abu2019mixhop}
\bibfield{author}{\bibinfo{person}{Sami Abu-El-Haija}, \bibinfo{person}{Bryan Perozzi}, \bibinfo{person}{Amol Kapoor}, \bibinfo{person}{Nazanin Alipourfard}, \bibinfo{person}{Kristina Lerman}, \bibinfo{person}{Hrayr Harutyunyan}, \bibinfo{person}{Greg Ver~Steeg}, {and} \bibinfo{person}{Aram Galstyan}.} \bibinfo{year}{2019}\natexlab{}.
\newblock \showarticletitle{Mixhop: Higher-order graph convolutional architectures via sparsified neighborhood mixing}. In \bibinfo{booktitle}{\emph{international conference on machine learning}}. PMLR, \bibinfo{pages}{21--29}.
\newblock


\bibitem[Bhojanapalli et~al\mbox{.}(2020)]%
        {srinadh2020lowrank}
\bibfield{author}{\bibinfo{person}{Srinadh Bhojanapalli}, \bibinfo{person}{Chulhee Yun}, \bibinfo{person}{Ankit~Singh Rawat}, \bibinfo{person}{Sashank~J. Reddi}, {and} \bibinfo{person}{Sanjiv Kumar}.} \bibinfo{year}{2020}\natexlab{}.
\newblock \showarticletitle{Low-Rank Bottleneck in Multi-head Attention Models}. In \bibinfo{booktitle}{\emph{International Conference on Machine Learning}}.
\newblock


\bibitem[Bojchevski et~al\mbox{.}(2020)]%
        {bojchevski2020scaling}
\bibfield{author}{\bibinfo{person}{Aleksandar Bojchevski}, \bibinfo{person}{Johannes Gasteiger}, \bibinfo{person}{Bryan Perozzi}, \bibinfo{person}{Amol Kapoor}, \bibinfo{person}{Martin Blais}, \bibinfo{person}{Benedek R{\'o}zemberczki}, \bibinfo{person}{Michal Lukasik}, {and} \bibinfo{person}{Stephan G{\"u}nnemann}.} \bibinfo{year}{2020}\natexlab{}.
\newblock \showarticletitle{Scaling graph neural networks with approximate pagerank}. In \bibinfo{booktitle}{\emph{Proceedings of the 26th ACM SIGKDD international conference on knowledge discovery \& data mining}}. \bibinfo{pages}{2464--2473}.
\newblock


\bibitem[Bronstein et~al\mbox{.}(2021)]%
        {bronstein2021geometric}
\bibfield{author}{\bibinfo{person}{Michael~M Bronstein}, \bibinfo{person}{Joan Bruna}, \bibinfo{person}{Taco Cohen}, {and} \bibinfo{person}{Petar Veli{\v{c}}kovi{\'c}}.} \bibinfo{year}{2021}\natexlab{}.
\newblock \showarticletitle{Geometric deep learning: Grids, groups, graphs, geodesics, and gauges}.
\newblock \bibinfo{journal}{\emph{arXiv preprint arXiv:2104.13478}} (\bibinfo{year}{2021}).
\newblock


\bibitem[Brown et~al\mbox{.}(2020)]%
        {NEURIPS2020_1457c0d6}
\bibfield{author}{\bibinfo{person}{Tom Brown}, \bibinfo{person}{Benjamin Mann}, \bibinfo{person}{Nick Ryder}, \bibinfo{person}{Melanie Subbiah}, \bibinfo{person}{Jared~D Kaplan}, \bibinfo{person}{Prafulla Dhariwal}, \bibinfo{person}{Arvind Neelakantan}, \bibinfo{person}{Pranav Shyam}, \bibinfo{person}{Girish Sastry}, \bibinfo{person}{Amanda Askell}, \bibinfo{person}{Sandhini Agarwal}, \bibinfo{person}{Ariel Herbert-Voss}, \bibinfo{person}{Gretchen Krueger}, \bibinfo{person}{Tom Henighan}, \bibinfo{person}{Rewon Child}, \bibinfo{person}{Aditya Ramesh}, \bibinfo{person}{Daniel Ziegler}, \bibinfo{person}{Jeffrey Wu}, \bibinfo{person}{Clemens Winter}, \bibinfo{person}{Chris Hesse}, \bibinfo{person}{Mark Chen}, \bibinfo{person}{Eric Sigler}, \bibinfo{person}{Mateusz Litwin}, \bibinfo{person}{Scott Gray}, \bibinfo{person}{Benjamin Chess}, \bibinfo{person}{Jack Clark}, \bibinfo{person}{Christopher Berner}, \bibinfo{person}{Sam McCandlish}, \bibinfo{person}{Alec Radford}, \bibinfo{person}{Ilya Sutskever}, {and}
  \bibinfo{person}{Dario Amodei}.} \bibinfo{year}{2020}\natexlab{}.
\newblock \showarticletitle{Language Models are Few-Shot Learners}. In \bibinfo{booktitle}{\emph{Advances in Neural Information Processing Systems}}. \bibinfo{pages}{1877--1901}.
\newblock


\bibitem[Chamberlain et~al\mbox{.}(2022)]%
        {chamberlain2022graph}
\bibfield{author}{\bibinfo{person}{Benjamin~Paul Chamberlain}, \bibinfo{person}{Sergey Shirobokov}, \bibinfo{person}{Emanuele Rossi}, \bibinfo{person}{Fabrizio Frasca}, \bibinfo{person}{Thomas Markovich}, \bibinfo{person}{Nils Hammerla}, \bibinfo{person}{Michael~M Bronstein}, {and} \bibinfo{person}{Max Hansmire}.} \bibinfo{year}{2022}\natexlab{}.
\newblock \showarticletitle{Graph neural networks for link prediction with subgraph sketching}.
\newblock \bibinfo{journal}{\emph{arXiv preprint arXiv:2209.15486}} (\bibinfo{year}{2022}).
\newblock


\bibitem[Chen et~al\mbox{.}(2022b)]%
        {chen2022structure}
\bibfield{author}{\bibinfo{person}{Dexiong Chen}, \bibinfo{person}{Leslie O’Bray}, {and} \bibinfo{person}{Karsten Borgwardt}.} \bibinfo{year}{2022}\natexlab{b}.
\newblock \showarticletitle{Structure-aware transformer for graph representation learning}. In \bibinfo{booktitle}{\emph{International conference on machine learning}}. PMLR, \bibinfo{pages}{3469--3489}.
\newblock


\bibitem[Chen et~al\mbox{.}(2022a)]%
        {chen2022nagphormer}
\bibfield{author}{\bibinfo{person}{Jinsong Chen}, \bibinfo{person}{Kaiyuan Gao}, \bibinfo{person}{Gaichao Li}, {and} \bibinfo{person}{Kun He}.} \bibinfo{year}{2022}\natexlab{a}.
\newblock \showarticletitle{NAGphormer: A tokenized graph transformer for node classification in large graphs}.
\newblock \bibinfo{journal}{\emph{arXiv preprint arXiv:2206.04910}} (\bibinfo{year}{2022}).
\newblock


\bibitem[Chen et~al\mbox{.}(2024)]%
        {chen2024hight}
\bibfield{author}{\bibinfo{person}{Yongqiang Chen}, \bibinfo{person}{Quanming Yao}, \bibinfo{person}{Juzheng Zhang}, \bibinfo{person}{James Cheng}, {and} \bibinfo{person}{Yatao Bian}.} \bibinfo{year}{2024}\natexlab{}.
\newblock \showarticletitle{Hight: Hierarchical graph tokenization for graph-language alignment}.
\newblock \bibinfo{journal}{\emph{arXiv preprint arXiv:2406.14021}} (\bibinfo{year}{2024}).
\newblock


\bibitem[Chien et~al\mbox{.}(2020)]%
        {chien2020adaptive}
\bibfield{author}{\bibinfo{person}{Eli Chien}, \bibinfo{person}{Jianhao Peng}, \bibinfo{person}{Pan Li}, {and} \bibinfo{person}{Olgica Milenkovic}.} \bibinfo{year}{2020}\natexlab{}.
\newblock \showarticletitle{Adaptive universal generalized pagerank graph neural network}.
\newblock \bibinfo{journal}{\emph{arXiv preprint arXiv:2006.07988}} (\bibinfo{year}{2020}).
\newblock


\bibitem[Deng et~al\mbox{.}(2024)]%
        {deng2024polynormer}
\bibfield{author}{\bibinfo{person}{Chenhui Deng}, \bibinfo{person}{Zichao Yue}, {and} \bibinfo{person}{Zhiru Zhang}.} \bibinfo{year}{2024}\natexlab{}.
\newblock \showarticletitle{Polynormer: Polynomial-expressive graph transformer in linear time}.
\newblock \bibinfo{journal}{\emph{arXiv preprint arXiv:2403.01232}} (\bibinfo{year}{2024}).
\newblock


\bibitem[Devlin et~al\mbox{.}(2019)]%
        {devlin-etal-2019-bert}
\bibfield{author}{\bibinfo{person}{Jacob Devlin}, \bibinfo{person}{Ming-Wei Chang}, \bibinfo{person}{Kenton Lee}, {and} \bibinfo{person}{Kristina Toutanova}.} \bibinfo{year}{2019}\natexlab{}.
\newblock \showarticletitle{{BERT}: Pre-training of Deep Bidirectional Transformers for Language Understanding}. In \bibinfo{booktitle}{\emph{Proceedings of the 2019 Conference of the North {A}merican Chapter of the Association for Computational Linguistics: Human Language Technologies}}. \bibinfo{pages}{4171--4186}.
\newblock


\bibitem[Ding et~al\mbox{.}(2021)]%
        {ding2021vqgnn}
\bibfield{author}{\bibinfo{person}{Mucong Ding}, \bibinfo{person}{Kezhi Kong}, \bibinfo{person}{Jingling Li}, \bibinfo{person}{Chen Zhu}, \bibinfo{person}{John~P Dickerson}, \bibinfo{person}{Furong Huang}, {and} \bibinfo{person}{Tom Goldstein}.} \bibinfo{year}{2021}\natexlab{}.
\newblock \showarticletitle{{VQ}-{GNN}: A Universal Framework to Scale up Graph Neural Networks using Vector Quantization}. In \bibinfo{booktitle}{\emph{Advances in Neural Information Processing Systems}}, \bibfield{editor}{\bibinfo{person}{A.~Beygelzimer}, \bibinfo{person}{Y.~Dauphin}, \bibinfo{person}{P.~Liang}, {and} \bibinfo{person}{J.~Wortman Vaughan}} (Eds.).
\newblock


\bibitem[Dosovitskiy et~al\mbox{.}(2021)]%
        {dosovitskiy2021an}
\bibfield{author}{\bibinfo{person}{Alexey Dosovitskiy}, \bibinfo{person}{Lucas Beyer}, \bibinfo{person}{Alexander Kolesnikov}, \bibinfo{person}{Dirk Weissenborn}, \bibinfo{person}{Xiaohua Zhai}, \bibinfo{person}{Thomas Unterthiner}, \bibinfo{person}{Mostafa Dehghani}, \bibinfo{person}{Matthias Minderer}, \bibinfo{person}{Georg Heigold}, \bibinfo{person}{Sylvain Gelly}, \bibinfo{person}{Jakob Uszkoreit}, {and} \bibinfo{person}{Neil Houlsby}.} \bibinfo{year}{2021}\natexlab{}.
\newblock \showarticletitle{An Image is Worth 16x16 Words: Transformers for Image Recognition at Scale}. In \bibinfo{booktitle}{\emph{International Conference on Learning Representations}}.
\newblock


\bibitem[Dwivedi and Bresson(2020)]%
        {dwivedi2020generalization}
\bibfield{author}{\bibinfo{person}{Vijay~Prakash Dwivedi} {and} \bibinfo{person}{Xavier Bresson}.} \bibinfo{year}{2020}\natexlab{}.
\newblock \showarticletitle{A generalization of transformer networks to graphs}.
\newblock \bibinfo{journal}{\emph{arXiv preprint arXiv:2012.09699}} (\bibinfo{year}{2020}).
\newblock


\bibitem[Dwivedi et~al\mbox{.}(2023)]%
        {dwivedi2023benchmarking}
\bibfield{author}{\bibinfo{person}{Vijay~Prakash Dwivedi}, \bibinfo{person}{Chaitanya~K Joshi}, \bibinfo{person}{Anh~Tuan Luu}, \bibinfo{person}{Thomas Laurent}, \bibinfo{person}{Yoshua Bengio}, {and} \bibinfo{person}{Xavier Bresson}.} \bibinfo{year}{2023}\natexlab{}.
\newblock \showarticletitle{Benchmarking graph neural networks}.
\newblock \bibinfo{journal}{\emph{Journal of Machine Learning Research}} \bibinfo{volume}{24}, \bibinfo{number}{43} (\bibinfo{year}{2023}), \bibinfo{pages}{1--48}.
\newblock


\bibitem[Fan et~al\mbox{.}(2025)]%
        {fan2025dual}
\bibfield{author}{\bibinfo{person}{Li Fan}, \bibinfo{person}{Menglin Kong}, \bibinfo{person}{Yang Xiang}, \bibinfo{person}{Chong Zhang}, {and} \bibinfo{person}{Chengtao Ji}.} \bibinfo{year}{2025}\natexlab{}.
\newblock \showarticletitle{Dual prototype attentive graph network for cross-market recommendation}. In \bibinfo{booktitle}{\emph{International Conference on Neural Information Processing}}. Springer, \bibinfo{pages}{517--531}.
\newblock


\bibitem[Fan et~al\mbox{.}(2019)]%
        {fan2019graph}
\bibfield{author}{\bibinfo{person}{Wenqi Fan}, \bibinfo{person}{Yao Ma}, \bibinfo{person}{Qing Li}, \bibinfo{person}{Yuan He}, \bibinfo{person}{Eric Zhao}, \bibinfo{person}{Jiliang Tang}, {and} \bibinfo{person}{Dawei Yin}.} \bibinfo{year}{2019}\natexlab{}.
\newblock \showarticletitle{Graph neural networks for social recommendation}. In \bibinfo{booktitle}{\emph{The world wide web conference}}. \bibinfo{pages}{417--426}.
\newblock


\bibitem[Galkin et~al\mbox{.}(2021)]%
        {galkin2021nodepiece}
\bibfield{author}{\bibinfo{person}{Mikhail Galkin}, \bibinfo{person}{Etienne Denis}, \bibinfo{person}{Jiapeng Wu}, {and} \bibinfo{person}{William~L Hamilton}.} \bibinfo{year}{2021}\natexlab{}.
\newblock \showarticletitle{Nodepiece: Compositional and parameter-efficient representations of large knowledge graphs}.
\newblock \bibinfo{journal}{\emph{arXiv preprint arXiv:2106.12144}} (\bibinfo{year}{2021}).
\newblock


\bibitem[Gasteiger et~al\mbox{.}(2018)]%
        {gasteiger2018predict}
\bibfield{author}{\bibinfo{person}{Johannes Gasteiger}, \bibinfo{person}{Aleksandar Bojchevski}, {and} \bibinfo{person}{Stephan G{\"u}nnemann}.} \bibinfo{year}{2018}\natexlab{}.
\newblock \showarticletitle{Predict then propagate: Graph neural networks meet personalized pagerank}.
\newblock \bibinfo{journal}{\emph{arXiv preprint arXiv:1810.05997}} (\bibinfo{year}{2018}).
\newblock


\bibitem[Guan et~al\mbox{.}(2025)]%
        {guan2025lihai}
\bibfield{author}{\bibinfo{person}{Yifan Guan}, \bibinfo{person}{Wei Wang}, \bibinfo{person}{Jianjun Chen}, \bibinfo{person}{Po Yang}, \bibinfo{person}{Jingzhou Xu}, {and} \bibinfo{person}{Jun Qi}.} \bibinfo{year}{2025}\natexlab{}.
\newblock \showarticletitle{A Survey of Multimodal Fusion for Alzheimer’s Disease Prediction: A New Taxonomy and Trends}.
\newblock \bibinfo{journal}{\emph{Information Fusion}} (\bibinfo{year}{2025}), \bibinfo{pages}{104098}.
\newblock
\showISSN{1566-2535}
\href{https://doi.org/10.1016/j.inffus.2025.104098}{doi:\nolinkurl{10.1016/j.inffus.2025.104098}}


\bibitem[Guo et~al\mbox{.}(2021)]%
        {guo2021graphcodebert}
\bibfield{author}{\bibinfo{person}{Daya Guo}, \bibinfo{person}{Shuo Ren}, \bibinfo{person}{Shuai Lu}, \bibinfo{person}{Zhangyin Feng}, \bibinfo{person}{Duyu Tang}, \bibinfo{person}{Shujie LIU}, \bibinfo{person}{Long Zhou}, \bibinfo{person}{Nan Duan}, \bibinfo{person}{Alexey Svyatkovskiy}, \bibinfo{person}{Shengyu Fu}, \bibinfo{person}{Michele Tufano}, \bibinfo{person}{Shao~Kun Deng}, \bibinfo{person}{Colin Clement}, \bibinfo{person}{Dawn Drain}, \bibinfo{person}{Neel Sundaresan}, \bibinfo{person}{Jian Yin}, \bibinfo{person}{Daxin Jiang}, {and} \bibinfo{person}{Ming Zhou}.} \bibinfo{year}{2021}\natexlab{}.
\newblock \showarticletitle{GraphCode{\{}BERT{\}}: Pre-training Code Representations with Data Flow}. In \bibinfo{booktitle}{\emph{International Conference on Learning Representations}}.
\newblock
\urldef\tempurl%
\url{https://openreview.net/forum?id=jLoC4ez43PZ}
\showURL{%
\tempurl}


\bibitem[Hamilton et~al\mbox{.}(2017)]%
        {hamilton2017inductive}
\bibfield{author}{\bibinfo{person}{Will Hamilton}, \bibinfo{person}{Zhitao Ying}, {and} \bibinfo{person}{Jure Leskovec}.} \bibinfo{year}{2017}\natexlab{}.
\newblock \showarticletitle{Inductive representation learning on large graphs}.
\newblock \bibinfo{journal}{\emph{Advances in neural information processing systems}}  \bibinfo{volume}{30} (\bibinfo{year}{2017}).
\newblock


\bibitem[Hou et~al\mbox{.}(2025)]%
        {hou2025heterogeneous}
\bibfield{author}{\bibinfo{person}{Hongru Hou}, \bibinfo{person}{Jiachen Sun}, \bibinfo{person}{Wenqing Lin}, \bibinfo{person}{Wendong Bi}, \bibinfo{person}{Xiangrong Wang}, {and} \bibinfo{person}{Deqing Yang}.} \bibinfo{year}{2025}\natexlab{}.
\newblock \showarticletitle{Heterogeneous Influence Maximization in User Recommendation}. In \bibinfo{booktitle}{\emph{Proceedings of the 34th ACM International Conference on Information and Knowledge Management}}. \bibinfo{pages}{5747--5754}.
\newblock


\bibitem[Hou et~al\mbox{.}(2023)]%
        {hou2023graphmae2}
\bibfield{author}{\bibinfo{person}{Zhenyu Hou}, \bibinfo{person}{Yufei He}, \bibinfo{person}{Yukuo Cen}, \bibinfo{person}{Xiao Liu}, \bibinfo{person}{Yuxiao Dong}, \bibinfo{person}{Evgeny Kharlamov}, {and} \bibinfo{person}{Jie Tang}.} \bibinfo{year}{2023}\natexlab{}.
\newblock \showarticletitle{Graphmae2: A decoding-enhanced masked self-supervised graph learner}. In \bibinfo{booktitle}{\emph{Proceedings of the ACM web conference 2023}}. \bibinfo{pages}{737--746}.
\newblock


\bibitem[Hou et~al\mbox{.}(2022)]%
        {hou2022graphmae}
\bibfield{author}{\bibinfo{person}{Zhenyu Hou}, \bibinfo{person}{Xiao Liu}, \bibinfo{person}{Yukuo Cen}, \bibinfo{person}{Yuxiao Dong}, \bibinfo{person}{Hongxia Yang}, \bibinfo{person}{Chunjie Wang}, {and} \bibinfo{person}{Jie Tang}.} \bibinfo{year}{2022}\natexlab{}.
\newblock \showarticletitle{Graphmae: Self-supervised masked graph autoencoders}. In \bibinfo{booktitle}{\emph{Proceedings of the 28th ACM SIGKDD conference on knowledge discovery and data mining}}. \bibinfo{pages}{594--604}.
\newblock


\bibitem[Hu et~al\mbox{.}(2025)]%
        {hu2025ids}
\bibfield{author}{\bibinfo{person}{Peiyu Hu}, \bibinfo{person}{Wayne Lu}, {and} \bibinfo{person}{Jia Wang}.} \bibinfo{year}{2025}\natexlab{}.
\newblock \showarticletitle{From ids to semantics: A generative framework for cross-domain recommendation with adaptive semantic tokenization}.
\newblock \bibinfo{journal}{\emph{arXiv preprint arXiv:2511.08006}} (\bibinfo{year}{2025}).
\newblock


\bibitem[Hu et~al\mbox{.}(2020)]%
        {qiu2023gptgnn}
\bibfield{author}{\bibinfo{person}{Ziniu Hu}, \bibinfo{person}{Yuxiao Dong}, \bibinfo{person}{Kuansan Wang}, \bibinfo{person}{Kai-Wei Chang}, {and} \bibinfo{person}{Yizhou Sun}.} \bibinfo{year}{2020}\natexlab{}.
\newblock \showarticletitle{GPT-GNN: Generative Pre-Training of Graph Neural Networks}. In \bibinfo{booktitle}{\emph{Proceedings of the 26th ACM SIGKDD Conference on Knowledge Discovery and Data Mining}}.
\newblock


\bibitem[Hussain et~al\mbox{.}(2022)]%
        {hussain2022global}
\bibfield{author}{\bibinfo{person}{Md~Shamim Hussain}, \bibinfo{person}{Mohammed~J Zaki}, {and} \bibinfo{person}{Dharmashankar Subramanian}.} \bibinfo{year}{2022}\natexlab{}.
\newblock \showarticletitle{Global self-attention as a replacement for graph convolution}. In \bibinfo{booktitle}{\emph{Proceedings of the 28th ACM SIGKDD conference on knowledge discovery and data mining}}. \bibinfo{pages}{655--665}.
\newblock


\bibitem[Jang et~al\mbox{.}(2016)]%
        {jang2016categorical}
\bibfield{author}{\bibinfo{person}{Eric Jang}, \bibinfo{person}{Shixiang Gu}, {and} \bibinfo{person}{Ben Poole}.} \bibinfo{year}{2016}\natexlab{}.
\newblock \showarticletitle{Categorical reparameterization with gumbel-softmax}.
\newblock \bibinfo{journal}{\emph{arXiv preprint arXiv:1611.01144}} (\bibinfo{year}{2016}).
\newblock


\bibitem[Jin et~al\mbox{.}(2024)]%
        {jin2024large}
\bibfield{author}{\bibinfo{person}{Bowen Jin}, \bibinfo{person}{Gang Liu}, \bibinfo{person}{Chi Han}, \bibinfo{person}{Meng Jiang}, \bibinfo{person}{Heng Ji}, {and} \bibinfo{person}{Jiawei Han}.} \bibinfo{year}{2024}\natexlab{}.
\newblock \showarticletitle{Large language models on graphs: A comprehensive survey}.
\newblock \bibinfo{journal}{\emph{IEEE Transactions on Knowledge and Data Engineering}} (\bibinfo{year}{2024}).
\newblock


\bibitem[Kim et~al\mbox{.}(2022)]%
        {kim2022pure}
\bibfield{author}{\bibinfo{person}{Jinwoo Kim}, \bibinfo{person}{Dat Nguyen}, \bibinfo{person}{Seonwoo Min}, \bibinfo{person}{Sungjun Cho}, \bibinfo{person}{Moontae Lee}, \bibinfo{person}{Honglak Lee}, {and} \bibinfo{person}{Seunghoon Hong}.} \bibinfo{year}{2022}\natexlab{}.
\newblock \showarticletitle{Pure transformers are powerful graph learners}.
\newblock \bibinfo{journal}{\emph{Advances in Neural Information Processing Systems}}  \bibinfo{volume}{35} (\bibinfo{year}{2022}), \bibinfo{pages}{14582--14595}.
\newblock


\bibitem[Kipf and Welling(2016a)]%
        {kipf2016gcn}
\bibfield{author}{\bibinfo{person}{Thomas~N Kipf} {and} \bibinfo{person}{Max Welling}.} \bibinfo{year}{2016}\natexlab{a}.
\newblock \showarticletitle{Semi-supervised classification with graph convolutional networks}.
\newblock \bibinfo{journal}{\emph{arXiv preprint arXiv:1609.02907}} (\bibinfo{year}{2016}).
\newblock


\bibitem[Kipf and Welling(2016b)]%
        {kipf2016gae}
\bibfield{author}{\bibinfo{person}{Thomas~N Kipf} {and} \bibinfo{person}{Max Welling}.} \bibinfo{year}{2016}\natexlab{b}.
\newblock \showarticletitle{Variational graph auto-encoders}.
\newblock \bibinfo{journal}{\emph{arXiv preprint arXiv:1611.07308}} (\bibinfo{year}{2016}).
\newblock


\bibitem[Kreuzer et~al\mbox{.}(2021)]%
        {kreuzer2021rethinking}
\bibfield{author}{\bibinfo{person}{Devin Kreuzer}, \bibinfo{person}{Dominique Beaini}, \bibinfo{person}{Will Hamilton}, \bibinfo{person}{Vincent L{\'e}tourneau}, {and} \bibinfo{person}{Prudencio Tossou}.} \bibinfo{year}{2021}\natexlab{}.
\newblock \showarticletitle{Rethinking graph transformers with spectral attention}.
\newblock \bibinfo{journal}{\emph{Advances in Neural Information Processing Systems}}  \bibinfo{volume}{34} (\bibinfo{year}{2021}), \bibinfo{pages}{21618--21629}.
\newblock


\bibitem[Lee et~al\mbox{.}(2022)]%
        {lee2022autoregressive}
\bibfield{author}{\bibinfo{person}{Doyup Lee}, \bibinfo{person}{Chiheon Kim}, \bibinfo{person}{Saehoon Kim}, \bibinfo{person}{Minsu Cho}, {and} \bibinfo{person}{Wook-Shin Han}.} \bibinfo{year}{2022}\natexlab{}.
\newblock \showarticletitle{Autoregressive image generation using residual quantization}. In \bibinfo{booktitle}{\emph{Proceedings of the IEEE/CVF conference on computer vision and pattern recognition}}. \bibinfo{pages}{11523--11532}.
\newblock


\bibitem[Li et~al\mbox{.}(2023)]%
        {li2023survey}
\bibfield{author}{\bibinfo{person}{Yuhan Li}, \bibinfo{person}{Zhixun Li}, \bibinfo{person}{Peisong Wang}, \bibinfo{person}{Jia Li}, \bibinfo{person}{Xiangguo Sun}, \bibinfo{person}{Hong Cheng}, {and} \bibinfo{person}{Jeffrey~Xu Yu}.} \bibinfo{year}{2023}\natexlab{}.
\newblock \showarticletitle{A survey of graph meets large language model: Progress and future directions}.
\newblock \bibinfo{journal}{\emph{arXiv preprint arXiv:2311.12399}} (\bibinfo{year}{2023}).
\newblock


\bibitem[Liu et~al\mbox{.}(2023)]%
        {liu2023gapformer}
\bibfield{author}{\bibinfo{person}{Chuang Liu}, \bibinfo{person}{Yibing Zhan}, \bibinfo{person}{Xueqi Ma}, \bibinfo{person}{Liang Ding}, \bibinfo{person}{Dapeng Tao}, \bibinfo{person}{Jia Wu}, {and} \bibinfo{person}{Wenbin Hu}.} \bibinfo{year}{2023}\natexlab{}.
\newblock \showarticletitle{Gapformer: Graph Transformer with Graph Pooling for Node Classification.}. In \bibinfo{booktitle}{\emph{IJCAI}}. \bibinfo{pages}{2196--2205}.
\newblock


\bibitem[Liu et~al\mbox{.}(2025)]%
        {liu2025graph}
\bibfield{author}{\bibinfo{person}{Jiawei Liu}, \bibinfo{person}{Cheng Yang}, \bibinfo{person}{Zhiyuan Lu}, \bibinfo{person}{Junze Chen}, \bibinfo{person}{Yibo Li}, \bibinfo{person}{Mengmei Zhang}, \bibinfo{person}{Ting Bai}, \bibinfo{person}{Yuan Fang}, \bibinfo{person}{Lichao Sun}, \bibinfo{person}{Philip~S Yu}, {et~al\mbox{.}}} \bibinfo{year}{2025}\natexlab{}.
\newblock \showarticletitle{Graph foundation models: Concepts, opportunities and challenges}.
\newblock \bibinfo{journal}{\emph{IEEE Transactions on Pattern Analysis and Machine Intelligence}} (\bibinfo{year}{2025}).
\newblock


\bibitem[Luo et~al\mbox{.}(2025)]%
        {luo2025node}
\bibfield{author}{\bibinfo{person}{Yuankai Luo}, \bibinfo{person}{Hongkang Li}, \bibinfo{person}{Qijiong Liu}, \bibinfo{person}{Lei Shi}, {and} \bibinfo{person}{Xiao-Ming Wu}.} \bibinfo{year}{2025}\natexlab{}.
\newblock \showarticletitle{Node Identifiers: Compact, Discrete Representations for Efficient Graph Learning}. In \bibinfo{booktitle}{\emph{The Thirteenth International Conference on Learning Representations}}.
\newblock
\urldef\tempurl%
\url{https://openreview.net/forum?id=t9lS1lX9FQ}
\showURL{%
\tempurl}


\bibitem[Luo et~al\mbox{.}(2024)]%
        {luo2024nid}
\bibfield{author}{\bibinfo{person}{Yuankai Luo}, \bibinfo{person}{Qijiong Liu}, \bibinfo{person}{Lei Shi}, {and} \bibinfo{person}{Xiao-Ming Wu}.} \bibinfo{year}{2024}\natexlab{}.
\newblock \showarticletitle{Structure-aware semantic node identifiers for learning on graphs}.
\newblock \bibinfo{journal}{\emph{arXiv e-prints}} (\bibinfo{year}{2024}), \bibinfo{pages}{arXiv--2405}.
\newblock


\bibitem[Ma et~al\mbox{.}(2023)]%
        {ma2023graph}
\bibfield{author}{\bibinfo{person}{Liheng Ma}, \bibinfo{person}{Chen Lin}, \bibinfo{person}{Derek Lim}, \bibinfo{person}{Adriana Romero-Soriano}, \bibinfo{person}{Puneet~K Dokania}, \bibinfo{person}{Mark Coates}, \bibinfo{person}{Philip Torr}, {and} \bibinfo{person}{Ser-Nam Lim}.} \bibinfo{year}{2023}\natexlab{}.
\newblock \showarticletitle{Graph inductive biases in transformers without message passing}. In \bibinfo{booktitle}{\emph{International Conference on Machine Learning}}. PMLR, \bibinfo{pages}{23321--23337}.
\newblock


\bibitem[Martins and Astudillo(2016)]%
        {martins2016softmax}
\bibfield{author}{\bibinfo{person}{Andre Martins} {and} \bibinfo{person}{Ramon Astudillo}.} \bibinfo{year}{2016}\natexlab{}.
\newblock \showarticletitle{From softmax to sparsemax: A sparse model of attention and multi-label classification}. In \bibinfo{booktitle}{\emph{International conference on machine learning}}. PMLR, \bibinfo{pages}{1614--1623}.
\newblock


\bibitem[Page et~al\mbox{.}(1999)]%
        {page1999pagerank}
\bibfield{author}{\bibinfo{person}{Lawrence Page}, \bibinfo{person}{Sergey Brin}, \bibinfo{person}{Rajeev Motwani}, {and} \bibinfo{person}{Terry Winograd}.} \bibinfo{year}{1999}\natexlab{}.
\newblock \bibinfo{booktitle}{\emph{The PageRank citation ranking: Bringing order to the web.}}
\newblock \bibinfo{type}{{T}echnical {R}eport}. \bibinfo{institution}{Stanford infolab}.
\newblock


\bibitem[Peng et~al\mbox{.}(2024)]%
        {peng2024chatgraph}
\bibfield{author}{\bibinfo{person}{Yun Peng}, \bibinfo{person}{Sen Lin}, \bibinfo{person}{Qian Chen}, \bibinfo{person}{Shaowei Wang}, \bibinfo{person}{Lyu Xu}, \bibinfo{person}{Xiaojun Ren}, \bibinfo{person}{Yafei Li}, {and} \bibinfo{person}{Jianliang Xu}.} \bibinfo{year}{2024}\natexlab{}.
\newblock \showarticletitle{ChatGraph: Chat with Your Graphs}. In \bibinfo{booktitle}{\emph{2024 IEEE 40th International Conference on Data Engineering (ICDE)}}. IEEE, \bibinfo{pages}{5445--5448}.
\newblock


\bibitem[Qiu et~al\mbox{.}(2025)]%
        {qiu2025gated}
\bibfield{author}{\bibinfo{person}{Zihan Qiu}, \bibinfo{person}{Zekun Wang}, \bibinfo{person}{Bo Zheng}, \bibinfo{person}{Zeyu Huang}, \bibinfo{person}{Kaiyue Wen}, \bibinfo{person}{Songlin Yang}, \bibinfo{person}{Rui Men}, \bibinfo{person}{Le Yu}, \bibinfo{person}{Fei Huang}, \bibinfo{person}{Suozhi Huang}, {et~al\mbox{.}}} \bibinfo{year}{2025}\natexlab{}.
\newblock \showarticletitle{Gated Attention for Large Language Models: Non-linearity, Sparsity, and Attention-Sink-Free}.
\newblock \bibinfo{journal}{\emph{arXiv preprint arXiv:2505.06708}} (\bibinfo{year}{2025}).
\newblock


\bibitem[Ramp{\'a}{\v{s}}ek et~al\mbox{.}(2022)]%
        {rampavsek2022recipe}
\bibfield{author}{\bibinfo{person}{Ladislav Ramp{\'a}{\v{s}}ek}, \bibinfo{person}{Michael Galkin}, \bibinfo{person}{Vijay~Prakash Dwivedi}, \bibinfo{person}{Anh~Tuan Luu}, \bibinfo{person}{Guy Wolf}, {and} \bibinfo{person}{Dominique Beaini}.} \bibinfo{year}{2022}\natexlab{}.
\newblock \showarticletitle{Recipe for a general, powerful, scalable graph transformer}.
\newblock \bibinfo{journal}{\emph{Advances in Neural Information Processing Systems}}  \bibinfo{volume}{35} (\bibinfo{year}{2022}), \bibinfo{pages}{14501--14515}.
\newblock


\bibitem[Scarselli et~al\mbox{.}(2008)]%
        {scarselli2008graph}
\bibfield{author}{\bibinfo{person}{Franco Scarselli}, \bibinfo{person}{Marco Gori}, \bibinfo{person}{Ah~Chung Tsoi}, \bibinfo{person}{Markus Hagenbuchner}, {and} \bibinfo{person}{Gabriele Monfardini}.} \bibinfo{year}{2008}\natexlab{}.
\newblock \showarticletitle{The graph neural network model}.
\newblock \bibinfo{journal}{\emph{IEEE transactions on neural networks}} \bibinfo{volume}{20}, \bibinfo{number}{1} (\bibinfo{year}{2008}), \bibinfo{pages}{61--80}.
\newblock


\bibitem[Shirzad et~al\mbox{.}(2023)]%
        {shirzad2023exphormer}
\bibfield{author}{\bibinfo{person}{Hamed Shirzad}, \bibinfo{person}{Ameya Velingker}, \bibinfo{person}{Balaji Venkatachalam}, \bibinfo{person}{Danica~J Sutherland}, {and} \bibinfo{person}{Ali~Kemal Sinop}.} \bibinfo{year}{2023}\natexlab{}.
\newblock \showarticletitle{Exphormer: Sparse transformers for graphs}. In \bibinfo{booktitle}{\emph{International Conference on Machine Learning}}. PMLR, \bibinfo{pages}{31613--31632}.
\newblock


\bibitem[Shomer et~al\mbox{.}(2024)]%
        {shomer2024lpformer}
\bibfield{author}{\bibinfo{person}{Harry Shomer}, \bibinfo{person}{Yao Ma}, \bibinfo{person}{Haitao Mao}, \bibinfo{person}{Juanhui Li}, \bibinfo{person}{Bo Wu}, {and} \bibinfo{person}{Jiliang Tang}.} \bibinfo{year}{2024}\natexlab{}.
\newblock \showarticletitle{LPFormer: An Adaptive Graph Transformer for Link Prediction}. In \bibinfo{booktitle}{\emph{Proceedings of the 30th ACM SIGKDD Conference on Knowledge Discovery and Data Mining}} (Barcelona, Spain) \emph{(\bibinfo{series}{KDD '24})}. \bibinfo{publisher}{Association for Computing Machinery}, \bibinfo{address}{New York, NY, USA}, \bibinfo{pages}{2686–2698}.
\newblock
\showISBNx{9798400704901}
\href{https://doi.org/10.1145/3637528.3672025}{doi:\nolinkurl{10.1145/3637528.3672025}}


\bibitem[Thakoor et~al\mbox{.}(2021)]%
        {thakoor2021large}
\bibfield{author}{\bibinfo{person}{Shantanu Thakoor}, \bibinfo{person}{Corentin Tallec}, \bibinfo{person}{Mohammad~Gheshlaghi Azar}, \bibinfo{person}{Mehdi Azabou}, \bibinfo{person}{Eva~L Dyer}, \bibinfo{person}{Remi Munos}, \bibinfo{person}{Petar Veli{\v{c}}kovi{\'c}}, {and} \bibinfo{person}{Michal Valko}.} \bibinfo{year}{2021}\natexlab{}.
\newblock \showarticletitle{Large-scale representation learning on graphs via bootstrapping}.
\newblock \bibinfo{journal}{\emph{arXiv preprint arXiv:2102.06514}} (\bibinfo{year}{2021}).
\newblock


\bibitem[Vaswani et~al\mbox{.}(2017)]%
        {vaswani2017attention}
\bibfield{author}{\bibinfo{person}{Ashish Vaswani}, \bibinfo{person}{Noam Shazeer}, \bibinfo{person}{Niki Parmar}, \bibinfo{person}{Jakob Uszkoreit}, \bibinfo{person}{Llion Jones}, \bibinfo{person}{Aidan~N Gomez}, \bibinfo{person}{{\L}ukasz Kaiser}, {and} \bibinfo{person}{Illia Polosukhin}.} \bibinfo{year}{2017}\natexlab{}.
\newblock \showarticletitle{Attention is all you need}.
\newblock \bibinfo{journal}{\emph{Advances in neural information processing systems}}  \bibinfo{volume}{30} (\bibinfo{year}{2017}).
\newblock


\bibitem[Velickovic et~al\mbox{.}(2017a)]%
        {velickovic2017graph}
\bibfield{author}{\bibinfo{person}{Petar Velickovic}, \bibinfo{person}{Guillem Cucurull}, \bibinfo{person}{Arantxa Casanova}, \bibinfo{person}{Adriana Romero}, \bibinfo{person}{Pietro Lio}, \bibinfo{person}{Yoshua Bengio}, {et~al\mbox{.}}} \bibinfo{year}{2017}\natexlab{a}.
\newblock \showarticletitle{Graph attention networks}.
\newblock \bibinfo{journal}{\emph{stat}} \bibinfo{volume}{1050}, \bibinfo{number}{20} (\bibinfo{year}{2017}), \bibinfo{pages}{10--48550}.
\newblock


\bibitem[Velickovic et~al\mbox{.}(2017b)]%
        {velickovic2017gat}
\bibfield{author}{\bibinfo{person}{Petar Velickovic}, \bibinfo{person}{Guillem Cucurull}, \bibinfo{person}{Arantxa Casanova}, \bibinfo{person}{Adriana Romero}, \bibinfo{person}{Pietro Lio}, \bibinfo{person}{Yoshua Bengio}, {et~al\mbox{.}}} \bibinfo{year}{2017}\natexlab{b}.
\newblock \showarticletitle{Graph attention networks}.
\newblock \bibinfo{journal}{\emph{stat}} \bibinfo{volume}{1050}, \bibinfo{number}{20} (\bibinfo{year}{2017}), \bibinfo{pages}{10--48550}.
\newblock


\bibitem[Veli{\v{c}}kovi{\'c} et~al\mbox{.}(2018)]%
        {velivckovic2018deep}
\bibfield{author}{\bibinfo{person}{Petar Veli{\v{c}}kovi{\'c}}, \bibinfo{person}{William Fedus}, \bibinfo{person}{William~L Hamilton}, \bibinfo{person}{Pietro Li{\`o}}, \bibinfo{person}{Yoshua Bengio}, {and} \bibinfo{person}{R~Devon Hjelm}.} \bibinfo{year}{2018}\natexlab{}.
\newblock \showarticletitle{Deep graph infomax}.
\newblock \bibinfo{journal}{\emph{arXiv preprint arXiv:1809.10341}} (\bibinfo{year}{2018}).
\newblock


\bibitem[Wang et~al\mbox{.}(2025a)]%
        {wang2025learning}
\bibfield{author}{\bibinfo{person}{Limei Wang}, \bibinfo{person}{Kaveh Hassani}, \bibinfo{person}{Si Zhang}, \bibinfo{person}{Dongqi Fu}, \bibinfo{person}{Baichuan Yuan}, \bibinfo{person}{Weilin Cong}, \bibinfo{person}{Zhigang Hua}, \bibinfo{person}{Hao Wu}, \bibinfo{person}{Ning Yao}, {and} \bibinfo{person}{Bo Long}.} \bibinfo{year}{2025}\natexlab{a}.
\newblock \showarticletitle{Learning Graph Quantized Tokenizers}. In \bibinfo{booktitle}{\emph{The Thirteenth International Conference on Learning Representations}}.
\newblock
\urldef\tempurl%
\url{https://openreview.net/forum?id=oYSsbY3G4o}
\showURL{%
\tempurl}


\bibitem[Wang et~al\mbox{.}(2023)]%
        {wang2023neural}
\bibfield{author}{\bibinfo{person}{Xiyuan Wang}, \bibinfo{person}{Haotong Yang}, {and} \bibinfo{person}{Muhan Zhang}.} \bibinfo{year}{2023}\natexlab{}.
\newblock \showarticletitle{Neural common neighbor with completion for link prediction}.
\newblock \bibinfo{journal}{\emph{arXiv preprint arXiv:2302.00890}} (\bibinfo{year}{2023}).
\newblock


\bibitem[Wang et~al\mbox{.}(2025b)]%
        {wang2025graph}
\bibfield{author}{\bibinfo{person}{Zehong Wang}, \bibinfo{person}{Zheyuan Liu}, \bibinfo{person}{Tianyi Ma}, \bibinfo{person}{Jiazheng Li}, \bibinfo{person}{Zheyuan Zhang}, \bibinfo{person}{Xingbo Fu}, \bibinfo{person}{Yiyang Li}, \bibinfo{person}{Zhengqing Yuan}, \bibinfo{person}{Wei Song}, \bibinfo{person}{Yijun Ma}, {et~al\mbox{.}}} \bibinfo{year}{2025}\natexlab{b}.
\newblock \showarticletitle{Graph Foundation Models: A Comprehensive Survey}.
\newblock \bibinfo{journal}{\emph{arXiv preprint arXiv:2505.15116}} (\bibinfo{year}{2025}).
\newblock


\bibitem[Wu et~al\mbox{.}(2022)]%
        {wu2022nodeformer}
\bibfield{author}{\bibinfo{person}{Qitian Wu}, \bibinfo{person}{Wentao Zhao}, \bibinfo{person}{Zenan Li}, \bibinfo{person}{David~P Wipf}, {and} \bibinfo{person}{Junchi Yan}.} \bibinfo{year}{2022}\natexlab{}.
\newblock \showarticletitle{Nodeformer: A scalable graph structure learning transformer for node classification}.
\newblock \bibinfo{journal}{\emph{Advances in Neural Information Processing Systems}}  \bibinfo{volume}{35} (\bibinfo{year}{2022}), \bibinfo{pages}{27387--27401}.
\newblock


\bibitem[Wu et~al\mbox{.}(2023)]%
        {wu2023sgformer}
\bibfield{author}{\bibinfo{person}{Qitian Wu}, \bibinfo{person}{Wentao Zhao}, \bibinfo{person}{Chenxiao Yang}, \bibinfo{person}{Hengrui Zhang}, \bibinfo{person}{Fan Nie}, \bibinfo{person}{Haitian Jiang}, \bibinfo{person}{Yatao Bian}, {and} \bibinfo{person}{Junchi Yan}.} \bibinfo{year}{2023}\natexlab{}.
\newblock \showarticletitle{Sgformer: Simplifying and empowering transformers for large-graph representations}.
\newblock \bibinfo{journal}{\emph{Advances in Neural Information Processing Systems}}  \bibinfo{volume}{36} (\bibinfo{year}{2023}), \bibinfo{pages}{64753--64773}.
\newblock


\bibitem[Wu et~al\mbox{.}(2020)]%
        {wu2020comprehensive}
\bibfield{author}{\bibinfo{person}{Zonghan Wu}, \bibinfo{person}{Shirui Pan}, \bibinfo{person}{Fengwen Chen}, \bibinfo{person}{Guodong Long}, \bibinfo{person}{Chengqi Zhang}, {and} \bibinfo{person}{Philip~S Yu}.} \bibinfo{year}{2020}\natexlab{}.
\newblock \showarticletitle{A comprehensive survey on graph neural networks}.
\newblock \bibinfo{journal}{\emph{IEEE transactions on neural networks and learning systems}} \bibinfo{volume}{32}, \bibinfo{number}{1} (\bibinfo{year}{2020}), \bibinfo{pages}{4--24}.
\newblock


\bibitem[Xia et~al\mbox{.}(2023)]%
        {xia2023molebert}
\bibfield{author}{\bibinfo{person}{Jun Xia}, \bibinfo{person}{Chengshuai Zhao}, \bibinfo{person}{Bozhen Hu}, \bibinfo{person}{Zhangyang Gao}, \bibinfo{person}{Cheng Tan}, \bibinfo{person}{Yue Liu}, \bibinfo{person}{Siyuan Li}, {and} \bibinfo{person}{Stan~Z. Li}.} \bibinfo{year}{2023}\natexlab{}.
\newblock \showarticletitle{Mole-{BERT}: Rethinking Pre-training Graph Neural Networks for Molecules}. In \bibinfo{booktitle}{\emph{The Eleventh International Conference on Learning Representations}}.
\newblock


\bibitem[Xia et~al\mbox{.}(2024)]%
        {xia2024opengraph}
\bibfield{author}{\bibinfo{person}{Lianghao Xia}, \bibinfo{person}{Ben Kao}, {and} \bibinfo{person}{Chao Huang}.} \bibinfo{year}{2024}\natexlab{}.
\newblock \showarticletitle{OpenGraph: Towards Open Graph Foundation Models}. In \bibinfo{booktitle}{\emph{Findings of the Association for Computational Linguistics: EMNLP 2024}}. \bibinfo{publisher}{Association for Computational Linguistics}, \bibinfo{address}{Miami, Florida, USA}, \bibinfo{pages}{2365--2379}.
\newblock


\bibitem[Xiang et~al\mbox{.}(2025)]%
        {xiang2025harnessing}
\bibfield{author}{\bibinfo{person}{Yang Xiang}, \bibinfo{person}{Li Fan}, \bibinfo{person}{Chenke Yin}, \bibinfo{person}{Menglin Kong}, {and} \bibinfo{person}{Chengtao Ji}.} \bibinfo{year}{2025}\natexlab{}.
\newblock \showarticletitle{Harnessing Light for Cold-Start Recommendations: Leveraging Epistemic Uncertainty to Enhance Performance in User-Item Interactions} \emph{(\bibinfo{series}{CIKM '25})}. \bibinfo{publisher}{Association for Computing Machinery}, \bibinfo{pages}{5361–5365}.
\newblock
\showISBNx{9798400720406}
\href{https://doi.org/10.1145/3746252.3760793}{doi:\nolinkurl{10.1145/3746252.3760793}}


\bibitem[Xiao et~al\mbox{.}(2024)]%
        {xiao2024efficient}
\bibfield{author}{\bibinfo{person}{Guangxuan Xiao}, \bibinfo{person}{Yuandong Tian}, \bibinfo{person}{Beidi Chen}, \bibinfo{person}{Song Han}, {and} \bibinfo{person}{Mike Lewis}.} \bibinfo{year}{2024}\natexlab{}.
\newblock \showarticletitle{Efficient Streaming Language Models with Attention Sinks}. In \bibinfo{booktitle}{\emph{International Conference on Representation Learning}}, \bibfield{editor}{\bibinfo{person}{B.~Kim}, \bibinfo{person}{Y.~Yue}, \bibinfo{person}{S.~Chaudhuri}, \bibinfo{person}{K.~Fragkiadaki}, \bibinfo{person}{M.~Khan}, {and} \bibinfo{person}{Y.~Sun}} (Eds.), Vol.~\bibinfo{volume}{2024}. \bibinfo{pages}{21875--21895}.
\newblock
\urldef\tempurl%
\url{https://proceedings.iclr.cc/paper_files/paper/2024/file/5e5fd18f863cbe6d8ae392a93fd271c9-Paper-Conference.pdf}
\showURL{%
\tempurl}


\bibitem[Xu et~al\mbox{.}(2018)]%
        {xu2018representation}
\bibfield{author}{\bibinfo{person}{Keyulu Xu}, \bibinfo{person}{Chengtao Li}, \bibinfo{person}{Yonglong Tian}, \bibinfo{person}{Tomohiro Sonobe}, \bibinfo{person}{Ken-ichi Kawarabayashi}, {and} \bibinfo{person}{Stefanie Jegelka}.} \bibinfo{year}{2018}\natexlab{}.
\newblock \showarticletitle{Representation learning on graphs with jumping knowledge networks}. In \bibinfo{booktitle}{\emph{International conference on machine learning}}. pmlr, \bibinfo{pages}{5453--5462}.
\newblock


\bibitem[Yang et~al\mbox{.}(2024)]%
        {yang2024vqgraph}
\bibfield{author}{\bibinfo{person}{Ling Yang}, \bibinfo{person}{Ye Tian}, \bibinfo{person}{Minkai Xu}, \bibinfo{person}{Zhongyi Liu}, \bibinfo{person}{Shenda Hong}, \bibinfo{person}{Wei Qu}, \bibinfo{person}{Wentao Zhang}, \bibinfo{person}{Bin CUI}, \bibinfo{person}{Muhan Zhang}, {and} \bibinfo{person}{Jure Leskovec}.} \bibinfo{year}{2024}\natexlab{}.
\newblock \showarticletitle{VQGraph: Rethinking Graph Representation Space for Bridging GNNs and MLPs}. In \bibinfo{booktitle}{\emph{International Conference on Learning Representations}}.
\newblock


\bibitem[Yin et~al\mbox{.}(2025)]%
        {yin2025uncertainty}
\bibfield{author}{\bibinfo{person}{Chenke Yin}, \bibinfo{person}{Li Fan}, \bibinfo{person}{Jia Wang}, \bibinfo{person}{Dongxiao Hu}, \bibinfo{person}{Haichao Zhang}, \bibinfo{person}{Chong Zhang}, {and} \bibinfo{person}{Yang Xiang}.} \bibinfo{year}{2025}\natexlab{}.
\newblock \showarticletitle{Uncertainty-Aware Semantic Decoding for LLM-Based Sequential Recommendation}.
\newblock \bibinfo{journal}{\emph{arXiv preprint arXiv:2508.07210}} (\bibinfo{year}{2025}).
\newblock


\bibitem[Ying et~al\mbox{.}(2021)]%
        {ying2021transformers}
\bibfield{author}{\bibinfo{person}{Chengxuan Ying}, \bibinfo{person}{Tianle Cai}, \bibinfo{person}{Shengjie Luo}, \bibinfo{person}{Shuxin Zheng}, \bibinfo{person}{Guolin Ke}, \bibinfo{person}{Di He}, \bibinfo{person}{Yanming Shen}, {and} \bibinfo{person}{Tie-Yan Liu}.} \bibinfo{year}{2021}\natexlab{}.
\newblock \showarticletitle{Do transformers really perform badly for graph representation?}
\newblock \bibinfo{journal}{\emph{Advances in neural information processing systems}}  \bibinfo{volume}{34} (\bibinfo{year}{2021}), \bibinfo{pages}{28877--28888}.
\newblock


\bibitem[You et~al\mbox{.}(2020)]%
        {you2020graph}
\bibfield{author}{\bibinfo{person}{Yuning You}, \bibinfo{person}{Tianlong Chen}, \bibinfo{person}{Yongduo Sui}, \bibinfo{person}{Ting Chen}, \bibinfo{person}{Zhangyang Wang}, {and} \bibinfo{person}{Yang Shen}.} \bibinfo{year}{2020}\natexlab{}.
\newblock \showarticletitle{Graph contrastive learning with augmentations}.
\newblock \bibinfo{journal}{\emph{Advances in neural information processing systems}}  \bibinfo{volume}{33} (\bibinfo{year}{2020}), \bibinfo{pages}{5812--5823}.
\newblock


\bibitem[Yun et~al\mbox{.}(2021)]%
        {yun2021neo}
\bibfield{author}{\bibinfo{person}{Seongjun Yun}, \bibinfo{person}{Seoyoon Kim}, \bibinfo{person}{Junhyun Lee}, \bibinfo{person}{Jaewoo Kang}, {and} \bibinfo{person}{Hyunwoo~J Kim}.} \bibinfo{year}{2021}\natexlab{}.
\newblock \showarticletitle{Neo-gnns: Neighborhood overlap-aware graph neural networks for link prediction}.
\newblock \bibinfo{journal}{\emph{Advances in Neural Information Processing Systems}}  \bibinfo{volume}{34} (\bibinfo{year}{2021}), \bibinfo{pages}{13683--13694}.
\newblock


\bibitem[Zeng et~al\mbox{.}(2025)]%
        {zeng2025numina}
\bibfield{author}{\bibinfo{person}{Changyu Zeng}, \bibinfo{person}{Yifan Wang}, \bibinfo{person}{Zimu Wang}, \bibinfo{person}{Wei Wang}, \bibinfo{person}{Zhengni Yang}, \bibinfo{person}{Muyi Bao}, \bibinfo{person}{Jiming Xiao}, \bibinfo{person}{Ahn Nguyen}, {and} \bibinfo{person}{Yutao Yue}.} \bibinfo{year}{2025}\natexlab{}.
\newblock \showarticletitle{NUMINA: A Natural Understanding Benchmark for Multi-dimensional Intelligence and Numerical Reasoning Abilities}.
\newblock \bibinfo{journal}{\emph{arXiv preprint arXiv:2509.16656}} (\bibinfo{year}{2025}).
\newblock


\bibitem[Zhang et~al\mbox{.}(2025)]%
        {zhang2025llm}
\bibfield{author}{\bibinfo{person}{Jian Zhang}, \bibinfo{person}{Junyi Guo}, \bibinfo{person}{Junyi Yuan}, \bibinfo{person}{Huanda Lu}, \bibinfo{person}{Yanlin Zhou}, \bibinfo{person}{Fangyu Wu}, \bibinfo{person}{Qiufeng Wang}, {and} \bibinfo{person}{Dongming Lu}.} \bibinfo{year}{2025}\natexlab{}.
\newblock \showarticletitle{LLM-Driven Completeness and Consistency Evaluation for Cultural Heritage Data Augmentation in Cross-Modal Retrieval}. In \bibinfo{booktitle}{\emph{Proceedings of the 2025 Conference on Empirical Methods in Natural Language Processing}}. \bibinfo{pages}{19418--19428}.
\newblock


\bibitem[Zhang and Chen(2018)]%
        {zhang2018link}
\bibfield{author}{\bibinfo{person}{Muhan Zhang} {and} \bibinfo{person}{Yixin Chen}.} \bibinfo{year}{2018}\natexlab{}.
\newblock \showarticletitle{Link prediction based on graph neural networks}.
\newblock \bibinfo{journal}{\emph{Advances in neural information processing systems}}  \bibinfo{volume}{31} (\bibinfo{year}{2018}).
\newblock


\bibitem[Zhu et~al\mbox{.}(2021)]%
        {zhu2021neural}
\bibfield{author}{\bibinfo{person}{Zhaocheng Zhu}, \bibinfo{person}{Zuobai Zhang}, \bibinfo{person}{Louis-Pascal Xhonneux}, {and} \bibinfo{person}{Jian Tang}.} \bibinfo{year}{2021}\natexlab{}.
\newblock \showarticletitle{Neural bellman-ford networks: A general graph neural network framework for link prediction}.
\newblock \bibinfo{journal}{\emph{Advances in neural information processing systems}}  \bibinfo{volume}{34} (\bibinfo{year}{2021}), \bibinfo{pages}{29476--29490}.
\newblock


\bibitem[Zhuang et~al\mbox{.}(2023)]%
        {zhuang2023imold}
\bibfield{author}{\bibinfo{person}{Xiang Zhuang}, \bibinfo{person}{Qiang Zhang}, \bibinfo{person}{Keyan Ding}, \bibinfo{person}{Yatao Bian}, \bibinfo{person}{Xiao Wang}, \bibinfo{person}{Jingsong Lv}, \bibinfo{person}{Hongyang Chen}, {and} \bibinfo{person}{Huajun Chen}.} \bibinfo{year}{2023}\natexlab{}.
\newblock \showarticletitle{Learning invariant molecular representation in latent discrete space}.
\newblock \bibinfo{journal}{\emph{Advances in Neural Information Processing Systems}}  \bibinfo{volume}{36} (\bibinfo{year}{2023}), \bibinfo{pages}{78435--78452}.
\newblock


\end{thebibliography}

\end{document}